\documentclass[journal=apchd5,manuscript=letter]{achemso}

\usepackage[version=3]{mhchem} 
\usepackage{graphicx}
\usepackage{dcolumn}
\usepackage{bm}
\usepackage{amsmath}
\usepackage{esint}
\PassOptionsToPackage{normalem}{ulem}
\usepackage{ulem}
\usepackage{soul,xcolor}
\usepackage{color}
\usepackage{comment}
\author{Alexandre Mayer}
\affiliation[Namur]{Laboratoire de Physique du Solide, University of Namur, Rue de Bruxelles 61, 5000 Namur, Belgium}
\altaffiliation{These authors contributed equally to this work.}

\author{Hai Bi}
\affiliation{John A. Paulson School of Engineering and Applied Sciences, Harvard University, 9 Oxford Street, Cambridge, MA 02138, United States of America}
\alsoaffiliation{Ji Hua Laboratory, Island Ring South Street, Foshan, Guangdong 528000, P. R. China}

\author{Sarah Griesse-Nascimento}
\affiliation{John A. Paulson School of Engineering and Applied Sciences, Harvard University, 9 Oxford Street, Cambridge, MA 02138, United States of America}

\author{Benoit Hackens}
\affiliation{IMCN/NAPS, Université catholique de Louvain (UCLouvain), 1348 Louvain-la-Neuve, Belgium}

\author{Jérome Loicq}
\affiliation[CSL]{Centre Spatial de Liège, Avenue du Pré-Aily, 4031 Angleur, Belgium}
\alsoaffiliation{Faculty of Aerospace Engineering, Delft University of Technology, Delft, 2629 HS, Netherlands}

\author{Eric Mazur}
\affiliation{John A. Paulson School of Engineering and Applied Sciences, Harvard University, 9 Oxford Street, Cambridge, MA 02138, United States of America}

\author{Olivier Deparis}
\affiliation[Namur]{Laboratoire de Physique du Solide, University of Namur, Rue de Bruxelles 61, 5000 Namur, Belgium}

\author{Micha\"el Lobet}
\affiliation [Namur]{Laboratoire de Physique du Solide, University of Namur, Rue de Bruxelles 61, 5000 Namur, Belgium}
\alsoaffiliation{Centre Spatial de Liège, Avenue du Pré-Aily, 4031 Angleur, Belgium}
\alsoaffiliation{John A. Paulson School of Engineering and Applied Sciences, Harvard University, 9 Oxford Street, Cambridge, MA 02138, United States of America}
\altaffiliation{These authors contributed equally to this work.}
\email{michael.lobet@unamur.be}

\title[Ultra-broadband perfect absorbers using plasmonic metamaterials]
  {Genetic-algorithm-aided ultra-broadband perfect absorbers using plasmonic metamaterials}

\abbreviations{IR,NMR,UV}
\keywords{Ultra-broadband absorption, plasmonic metamaterials, genetic algorithm}

\begin{document}







\begin{abstract}

Complete absorption of electromagnetic waves is paramount in today's applications, ranging from photovoltaics to cross-talk prevention into sensitive devices. 
In this context, we use a genetic algorithm (GA) strategy to optimize absorption properties of
periodic arrays of truncated square-based pyramids made of alternating stacks of metal/dielectric layers. We target ultra-broadband quasi-perfect absorption of normally incident electromagnetic radiations in the visible and near-infrared ranges (wavelength comprised between 420 and 1600 nm).
We compare the results one can obtain by considering one, two or three stacks of 
 either Ni, Ti, Al, Cr, Ag, Cu, Au or W for the metal, and poly(methyl methacrylate) 
 (PMMA) for the dielectric. More than $10^{17}$ configurations of geometrical parameters are explored and reduced to a few optimal ones. This extensive study shows that Ni/PMMA, Ti/PMMA, Cr/PMMA and W/PMMA provide high-quality solutions with an integrated absorptance higher 
 than 99\% over the considered wavelength range, when considering realistic implementation of these ultra-broadband perfect electromagnetic absorbers. 
 Robustness of optimal solutions with respect to geometrical parameters is investigated and local absorption maps are provided.
 Moreover, we confirm that these optimal solutions maintain quasi-perfect broadband absorption properties 
 over a broad angular range when changing the inclination of the incident radiation.
 The study also reveals that noble metals (Au, Ag, Cu) do not provide the highest performance for the present application.
\end{abstract}


\section{Introduction}

Since its theoretical introduction in 1860 by G. Kirchhoff, the black body concept played a seminal role in the history of quantum mechanics and modern physics.\cite{Planck_1914} It originally referred to an \textit{idealized physical body of infinitely small thickness, that completely absorbs all incident rays, 
and neither reflects nor transmits any}.\cite{black_body}
The modern acceptance of the term does not include the infinitely small thickness anymore but still preserves the requirement to absorb all incident electromagnetic radiation regardless of the wavelength, the polarization or the angle of incidence of the incoming radiation. First experimental realizations of black bodies at the end of the 19th century/ dawn of the 20th century consisted of metallic boxes with its interior walls blackened by mixed chromium, nickel and cobalt oxides.
\cite{Lummer_1898,Lummer_1901,Mehra_2000}
The quest for a Perfect Electromagnetic Absorber (PEA), i.e. the materialization of the idealized black body, continued over the years for example due to the recent need of efficient solar energy harnessing or camouflage solutions, the development of photothermal detectors or in order to prevent crosstalk in nanoscale opto-electronics and quantum technologies.\cite{Barends_2011,Corcoles_2011} Among efficient solutions, carbon black and carbon nanotubes materials provide ultra-broadband PEA with 98-99 $\%$  absorption from UV to far infrared\cite{Mizuno_2009} while nickel-phosphorus alloy reach 96 \% on the $5-9 \mu m$ range.\cite{Ishii_2003}

At the dawn of the present century, metamaterials and plasmonic materials provided new opportunities in order to mold the flow of light at the nanoscale and drastically enhance light-matter interactions \cite{Maier_2007,Engheta2006,Joannopoulos2008,Lodahl2015}. Negative refraction, cloaking, superlensing, near-zero refractive index, surface enhanced Raman scattering, high energy concentrations at metal-dielectric interfaces are some examples of current interest among those hot topics in photonics.\cite{Pendry2000,Pendry2006,Liberal_2017}
Nevertheless, losses due to metallic components are an important drawback limiting current applications.\cite{Khurgin_2015}
This drawback was turned into an advantage by Landy \textit{et al}. who designed the first metamaterial PEA by using metallic resonators operating at a single wavelength.\cite{Landy_2008} The metamaterial approach was successfully applied during the last decade in order to tame the black body radiation and provide efficient PEA over a broad range of frequencies.\cite{Hedayati_2011,Hedayati_2012,Cui_2011,Cui_2012,Zhu_2012,Wang_2012,Ye_2010,Cui_2012b,Ding_2012,Argyropoulos_2013,Qiuqun_2013,Zhou_2021,Dixon_2020,Hajian19,Huang_2016,Ziegler2020}
Some proposed theoretical structures are nevertheless facing some technological difficulties due to the current limitations in resolution and complexity of the structures, especially in the visible range.
The field of plasmonic metamaterials PEA is now mature enough to enter a novel phase of \textit{in silico} modeling in order to develop more realistic 
PEA,\cite{Mayer_SPIE2018,Mayer_SPIE2020,Liu_2020} while preserving the ultra-broadband character of the PEA as well as its low angular dependency.

Here, we report on ultra-broadband metamaterial PEA using periodic stacks of square metal/dielectric layers arranged in a pyramidal way (see Fig. \ref{figure1}). We investigate among twenty-four possible PEA which are realistic from an experimental point of view and based on eight common metals arranged either in one, two or three stacks of metal/dielectric layers. The fact that the number of layers is limited to a few ones, while preserving the performances, 
is attractive in view of the tradeoff between fabrication complexity and broad absorption properties. 
It provides realistic perspectives for device fabrication with current technology.  
No less than $10^{17}$ configurations are explored by varying different experimental parameters such as the lateral size of the metal/dielectric layers, the thickness of the dielectric layer or the lateral periodicity between different individual structures. The search for the optimal configuration, i.e. the PEA 
with the largest absoptance of electromagnetic (EM) radiations over the visible to mid-infrared (MIR) range, is realized using a Genetic Algorithm (GA) strategy. Optimal configurations are reported as well as their absorptance spectra. Moreover, fields maps and angular dependency of the optimal PEA are discussed. The latter is primordial for several applications of black coatings such as cross-talk prevention in qubits.\cite{Barends_2011,Corcoles_2011} or stray light reduction\cite{Gong_2021,Fest_2013}.

\section{Design}
 
Energy conservation imposes that reflectance $R$, transmittance $T$ and absorptance $A$ are related through
\begin{equation}
 R(\lambda,\theta)+T(\lambda,\theta)+A(\lambda,\theta)=1
 \label{Energy conservation}
\end{equation}
with $\lambda$ being the wavelength of the incident EM radiation and $\theta$ the angle of incidence. Maximizing absorption requires suppressing both transmission and reflection at the same time. The PEA can be made opaque, i.e. no transmission, by adding a metallic layer over the substrate. In order to lower the reflection, the PEA needs to be impedance matched. Surface corrugations in a tapered way provide this gradual transition between the refractive indices of both incidence and substrate media. This anti-reflection strategy is well-known both in Nature  or in transparent coatings \cite{Clapham_1973,Deparis_2009}. The previous approach leads to suppressed transmission ($T=0$) and drastically reduced reflection ($R \to 0$).  To minimize reflection and fully absorb the incident radiation in order to obtain a PEA, the energy transported by the radiation has to be dissipated through the excitation of eigenmodes of the structure. 
This can be done for example by exciting localized surface plasmons (LSP) at the interfaces of metal/dielectric square resonators \cite{Prodan_2003,Christ_2006,Liu_2007,Pu_2012}.
The wavelength associated with these plasmonic resonances is proportional to the lateral dimensions of the metallic 
layers, conferring the desired broadband character once metallic resonators of varied sizes are stacked. 
Following this approach, truncated square-based pyramids made 
of twenty stacks of Au/Ge layers lead to an integrated absorptance of 98\% of normally incident radiations over a 0.2-5.8 $\mu$m wavelength
range\cite{Lobet_2014,Lobet2014bis}. However, fabrication of such elaborated structures, with nanometer resolution and high number of layers 
is currently too demanding in terms of time and resources.

\begin{figure}[t]
 \begin{center}
  \begin{tabular}{c}
   \includegraphics[width=12cm]{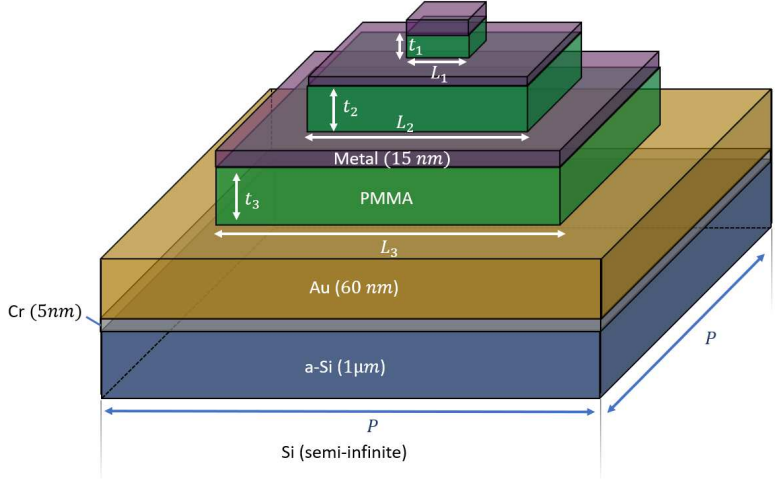}
  \end{tabular}
 \end{center}
 \caption{\label{figure1}Perfect Electromagnetic Absorber using square-based truncated pyramid made of $N=1,2$ or $3$ stacks of metal/PMMA layers. PMMA thickness $t_i$ and size $L_i$ of each layer is optimized using a Genetic Algorithm strategy, as well as the period $P$ of the PEA. The support of the pyramid consists
 of uniform layers of Au (60 nm), Cr (5 nm) and a-Si (1 micron); they rely on an infinite substrate of Si ($\varepsilon = 16$).}
\end{figure}

There is thus a need for simplified structures to reach ultra-broadband PEA that are tractable experimentally. Therefore, we consider here a simpler PEA 
consisting of one, two or three stacks of metal/dielectric layers (see Fig. \ref{figure1}).\cite{Mayer_SPIE2018,Mayer_SPIE2020}.
We set the thickness of all metal layers within the pyramid stack to 15 nm,\cite{Lobet_2014} which is smaller than the skin depth of the metals considered
over the targetted radiation wavelength range, i.e. from 420 to 1600 nm here. It allows therefore a coupling between the surface plasmons at the two sides of each metallic layer. Moreover, we consider poly(methyl methacrylate) (PMMA) as dielectric \cite{Beadie_2015,Mayer_SPIE2018,Mayer_SPIE2020}.
PMMA can be easily deposited using spin-coating techniques and its thickness can be simply varied by changing PMMA/solvent relative concentration in solution, 
or spinning speed for example.

A vast range of metals are considered in the present study, namely Ni, Ti, Al, Cr, Ag, Cu, Au or W.\cite{Johnson_1972,Johnson_1974,Rakic_1995,Rakic_1998}
Those metals correspond to widely used materials in current nanophotonics applications, and are readily deposited to form few nm-thick films through 
physical deposition techniques.
The thickness $t_i$ of each dielectric layer and the lateral dimension $L_i$ of each metal/dielectric stack are adjustable parameters in the present 
GA-aided approach. The lateral periodicity $P$ of the system also remains an adjustable parameter.
This relaxation of the PEA dimensions, as compared with previous studies\cite{Lobet_2014} where thickness was the same for all metallic layers 
and for all dielectric layers, provides the extra degrees of freedom that are necessary to maintain 
high performances while reducing the number of metallic layers.

The role of these different parameters can be understood as follows. The lateral dimensions $L_i$ of the metallic resonators essentially control the
plasmonic resonances of the system. The thicknesses $t_i$ of the dielectric layers control the vertical coupling between these resonators
while the periodicity $P$ finally controls the lateral coupling between the pyramids \cite{Lobet_2014}. The detailed variation range of 
these parameters are provided in SI. 
Furthermore, the PEA stands on a flat 60 nm-thick gold layer that blocks the transmission of 
the considered incident radiations and reflects any EM radiation not absorbed in the pyramids. This gold layer stands on a 5-nm thick 
chromium (Cr) adhesion layer followed by a 1 µm-thick amorphous silicon (a-Si) layer. We consider finally a semi-infinite Si 
substrate ($\varepsilon=16$). This choice of a silicon substrate has no impact on our results since no radiation will actually 
reach this region.

\section{Optimization schemes}

For a system made of $N$ stacks of metal/dielectric layers, we consequently have a total of $2N+1$ parameters to consider
(i.e., $L_i$, $t_i$ and $P$ for $i\in [1,N]$). 
Different methods are used in optics to determine optimal parameter combinations.
The systematic evaluation of a whole grid of possible parameter combinations is generally limited to two or three parameters. 
Local optimization methods such as the (quasi-)Newton method or gradient descent are more efficient
in terms of the required number of function evaluations.\cite{Dennis_1996} They converge however generally to the first 
local optimum encountered.
Global optimization methods are preferable in this respect since they provide a wider exploration of the parameter space,
which can eventually lead to higher-quality solutions (ideally the best-possible solution).
Genetic Algorithms (GA),\cite{Goldberg_1989,Haupt_2007,Eiben_2007}
Particle Swarm Optimization (PSO)\cite{Kennedy_1995,Shi_1998,Bonyadi_2017} and Ant Colony Optimization (ACO)\cite{Dorigo_2004}
are popular global optimization methods
that rely on a collective exploration of the parameter space. The population dynamics of these algorithms 
is inspired by the principles of natural selection for GA, birds flocks dynamics for PSO or 
pheromone-guided exploration for ACO. These methods can easily escape local optima and find 
higher-quality solutions. They do not require the calculation of gradients. 
They are intrinsically suited to a parallel computing, since the fitness of each individual in the population 
can be evaluated independently.
Machine Learning (ML) techniques such as deep neural networks or autoencoders can help the search by learning representations
of the function to optimize\cite{Gaier_2020}.

\begin{figure}[t]
 \begin{center}
  \begin{tabular}{c}
   \includegraphics[width=10cm]{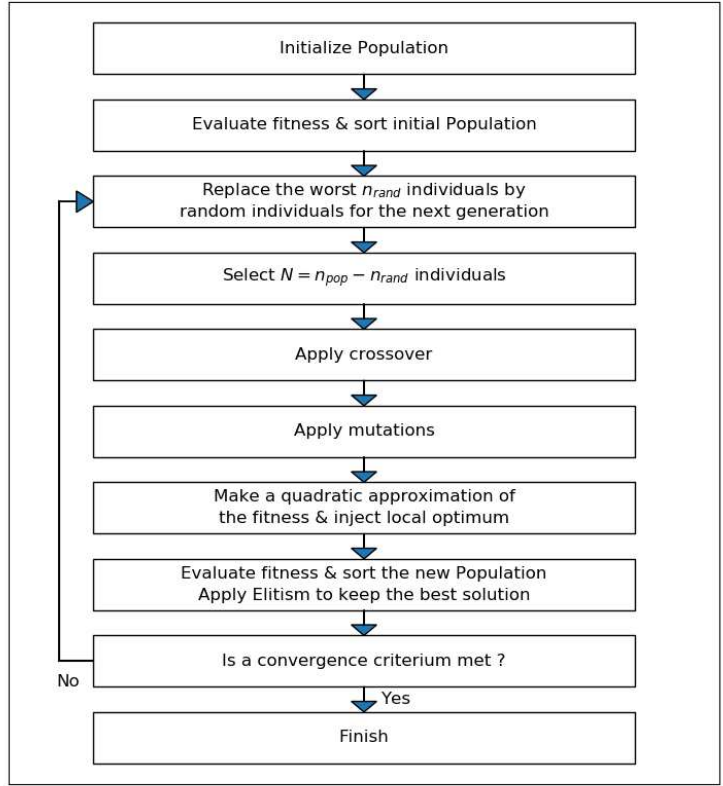}
  \end{tabular}
 \end{center}
 \caption{\label{figure2}Workflow of the Genetic Algorithm.}
\end{figure}

We use in this work a home-made genetic algorithm  (see workflow in Fig. \ref{figure2} and the 
Supplementary Material for a detailed description) to address this optimization problem and 
establish realistic structures that consist of only one, two or three stacks of 
metal/dielectric layers. We consider in particular poly(methyl methacrylate) (PMMA) for the dielectric and either Ni, Ti, Cr, Al, Ag, Cu, Au 
or W for the metal (eight possible materials). This makes a total of 24=3$\times$8 structures (1 to 3 layers, 8 metals)
that are optimized with the objective to maximize 
the absorption of normally incident radiations with wavelengths comprised between 420 and 1600 nm.

The objective function to be maximized (also called fitness or figure of merit) will be the integrated 
absorptance for normally incident radiations with wavelengths comprised between 420 and 1600 nm.
It is formally defined by
$\eta (\%) = 100 \times \frac{\int_{\lambda_{\rm min}}^{\lambda_{\rm max}} A(\lambda) d\lambda}{\lambda_{\rm max}-\lambda_{\rm min}}$,
where $A(\lambda)$ refers to the absorptance of normally incident radiations at the wavelength $\lambda$, 
$\lambda_{\rm min}$=420 nm and $\lambda_{\rm max}$=1600 nm. This is the quantity that the Genetic Algorithm seeks at 
maximizing by exploring the geometrical parameters of the structure considered.
It relies on an homemade Rigorous Coupled Waves Analysis (RCWA) code that solves Maxwell's equations exactly for stratified periodic systems.\cite{Moharam_1981,Deparis_2009}
A plane wave (PW) expansion of the electric permittivity of the PEA is made, in which 
the number of Fourier components is the key parameter to insure numerical convergence.


\section{Results and discussion}

\begin{figure}[t]
 \begin{center}
   \includegraphics[width=10cm]{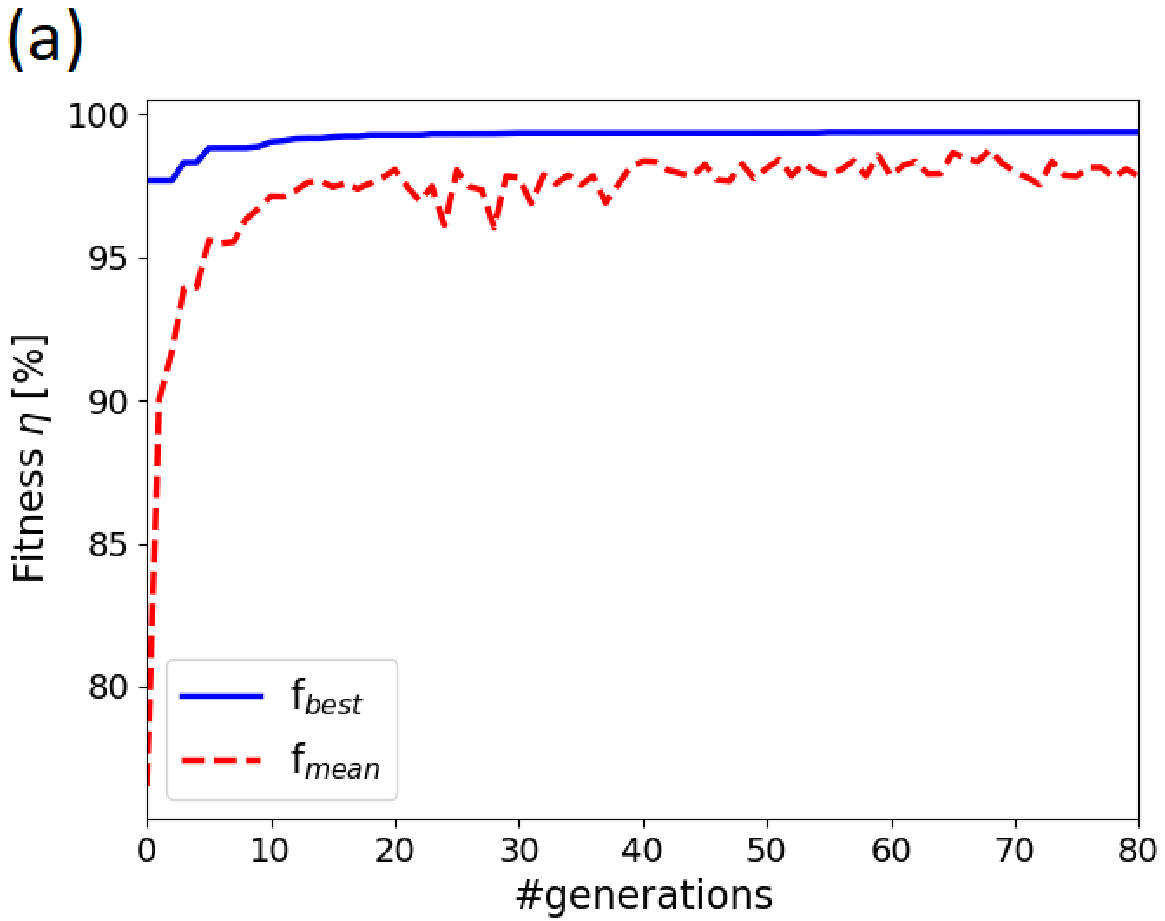}
 \end{center}
 \begin{center}
   \includegraphics[width=10cm]{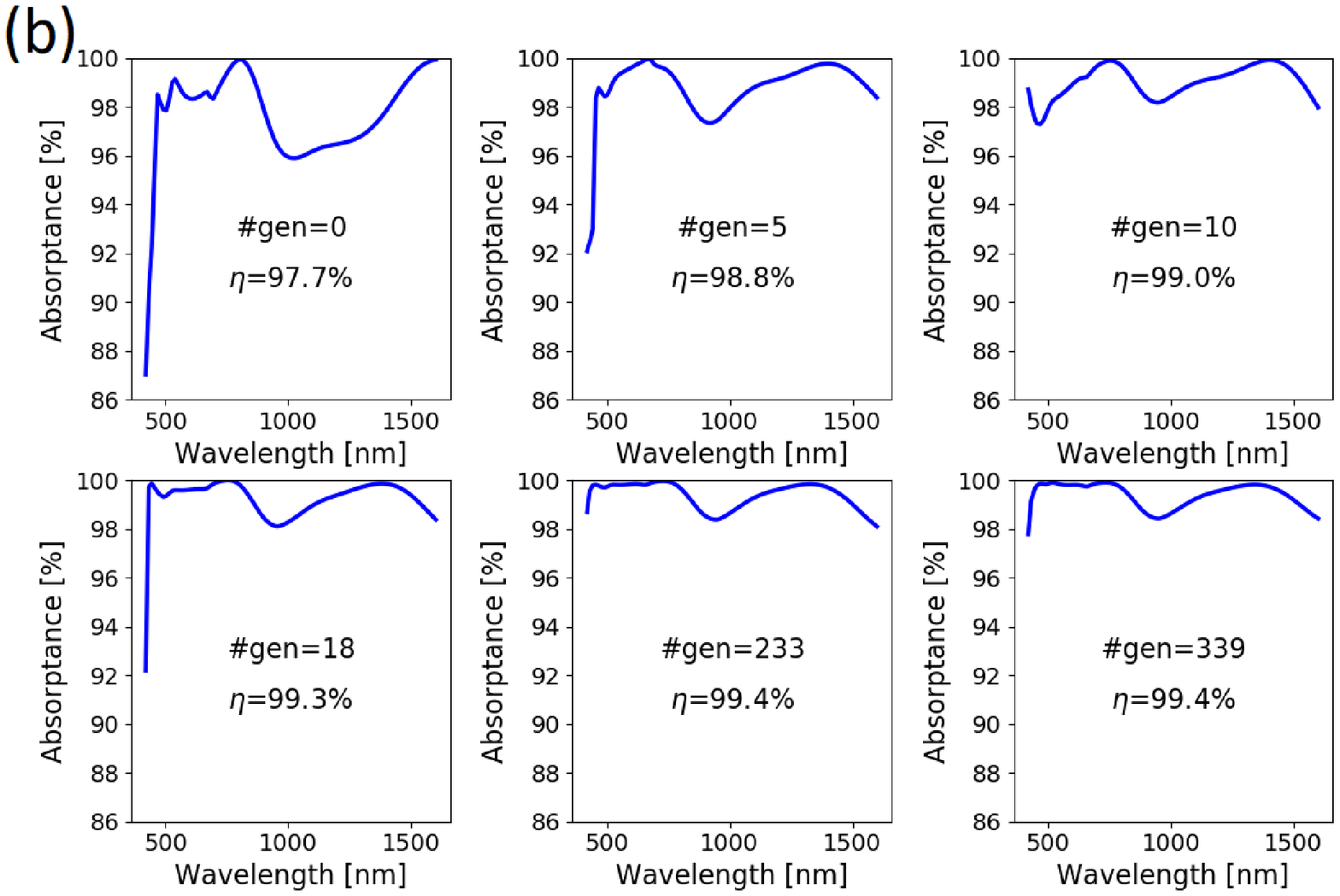}
 \end{center}
 \caption{\label{figure3} (a) best fitness (solid) and mean fitness (dashed) in the first 80 generations 
 when optimizing $N=3$ stacks of W/PMMA. (b) Absorptance spectrum for the best solution 
 obtained after 0, 5, 10, 18, 233 and 339 generations. The genetic algorithm was run with a population $n_{\rm pop}$
 of 50 individuals. The final solution has a fitness $\eta$ of 99.4\% when using $11\times 11$ PW in the 
 RCWA calculations.}
\end{figure}

Once the GA started, one can follow generation after generation the fitness of the best individual in the population as well as the mean fitness in the population. The best fitness usually increases rapidly in the first generations as better solutions are rapidly detected. Progress becomes typically slower after these first generations as the algorithm must either escape a local optimum to detect a radically different solution (exploration) or refine the best solution found so far to finalize the 
optimization (exploitation). A good optimization algorithm must actually find a sound balance between these two aspects.
The mean fitness of the population follows the best fitness, to a degree that depends on the convergence of the population
to the best individual, convergence measured by the genetic similarity $s$ defined in SM.
Fig. \ref{figure3} shows the best fitness and the mean fitness achieved in the first 80 generations, when optimizing $N=3$ stacks of W/PMMA. All optimizations are actually carried out by using $11\times 11$ plane waves in the RCWA simulations.
The figure also shows the absorptance spectrum for the best solutions found by the genetic algorithm after 0, 5, 10, 18, 233
and 339 generations. High-quality solutions are established rapidly by the GA in the first generations ($\eta=97.7\%$ at generation 0, 98.8\% at generation 5, 99.0\% at generation 10 and 99.3\% at generation 18). Qualitative improvements become then slower. The final solution ($\eta$=99.4\%) is found after 339 generations in this case.

\begin{figure}[t]
 \begin{center}
  \includegraphics[width=10cm]{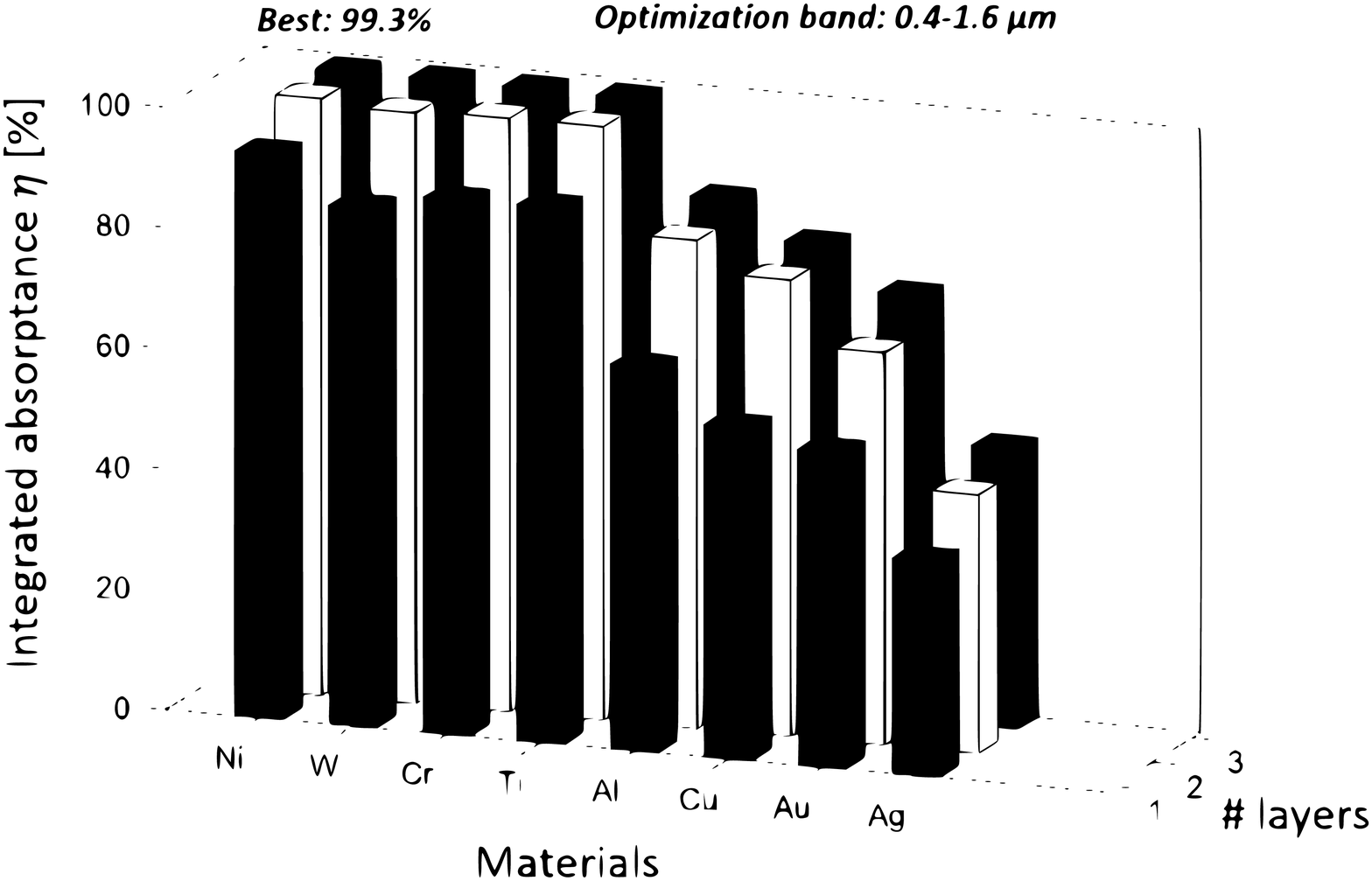}
 \end{center}
 \caption{\label{figure4}Figure of merit (integrated absorptance) $\eta$ achieved for each structure considered. These results correspond to
          RCWA calculations with $21 \times 21$ plane waves.}
\end{figure}

Fig. \ref{figure4} shows the best fitness (integrated absorptance $\eta$) achieved for the different structures considered. These results correspond to $N=1$, 2 and 3 stacks of metal/dielectric layers and the eight different metals considered in this study. They correspond to the $3\times 8=24$ systems actually optimized by the Genetic Algorithm (the optimization relies on RCWA calculations with $11 \times 11$ plane waves; the final solutions are confirmed by considering $21\times 21$ plane waves). This extensive study reveals that Ni, W, Cr or Ti represent the best materials for this application. It is noticeable that integrated absorptance values above 86.1\% are achieved for the $3 \times 4=12$ optimized structures that correspond to
these four metals. The best structures, which consist of $N=3$ stacks of Ni, W, Cr or Ti/PMMA layers, have all an integrated absorptance above 99\%. Those results are, to the best of our knowledge, the best PEA proposed in recent literature over such an extended wavelength range in the visible and NIR. 
The geometrical parameters and the figure of merit obtained for the four best metals selected (Ni, W, Cr and Ti) are given in Table \ref{table1}. The results that 
correspond to the other four metals (Al, Cu, Au and Ag) are given in Table \ref{table2}.  They will be discarded for this PEA application, essentially because of a lack of robustness of these solutions within the parameter space, i.e. sensitivity to experimental deviations,
in addition to lower integrated absorptance values.  It should be noted that three out of those four less convincing are noble metals (Cu, Au, Ag) and are usually considered as low-loss metals for plasmonic applications.\cite{Khurgin_2015,Khurgin_2017}
Considering that the choice of noble metals (Au, Ag, Cu) impacts on fabrication costs, it is interesting to note that they do not provide the highest performance for the present application.

\begin{table}[t]
 \vspace{0.4cm}
 {\scriptsize
 \begin{center}
  Ni / PMMA \vskip 7pt
  \begin{tabular}{|l|c|c|c|c|c|c|c||c|c|}
  \hline
     & $L_1$ (nm) & $t_1$ (nm) & $L_2$ (nm) & $t_2$ (nm) & $L_3$ (nm) & $t_3$ (nm) & $P$ (nm) & $\eta_{11\times 11{\rm PW}}$ & $\eta_{21\times 21{\rm PW}}$ \cr
  \hline
  \hline
   $N=1$ stack & {201} & {117} & - & - & - & - & {286} & {95.0\%} & {93.8\%} \cr
  \hline
  \hline
   $N=2$ stacks & {132} & {123} & {227} & {107} & - & - & {287} & {99.4\%} & {99.0\%} \cr
  \hline
  \hline
   $N=3$ stacks & {149} & {131} & {268} & {124} & {369} & {101} & {412} & {99.8\%} & {\bf 99.3\%} \cr
  \hline
  \end{tabular}
 \end{center}
 \begin{center}
  W / PMMA\vskip 7pt
  \begin{tabular}{|l|c|c|c|c|c|c|c||c|c|}
  \hline
     & $L_1$ (nm) & $t_1$ (nm) & $L_2$ (nm) & $t_2$ (nm) & $L_3$ (nm) & $t_3$ (nm) & $P$ (nm) & $\eta_{11\times 11{\rm PW}}$ & $\eta_{21\times 21{\rm PW}}$ \cr
  \hline
  \hline
   $N=1$ stack & {250} & {115} & - & - & - & - & {333} & {85.8\%} & {86.1\%} \cr
  \hline
  \hline
   $N=2$ stacks & {222} & {127} & {368} & {119} & - & - & {419} & {98.0\%} & {98.0\%} \cr
  \hline
  \hline
   $N=3$ stacks & {152} & {117} & {266} & {122} & {379} & {125} & {419} & {99.4\%} & {\bf 99.2\%} \cr
  \hline
  \end{tabular}
 \end{center}
 \begin{center}
  Cr / PMMA\vskip 7pt
  \begin{tabular}{|l|c|c|c|c|c|c|c||c|c|}
  \hline
     & $L_1$ (nm) & $t_1$ (nm) & $L_2$ (nm) & $t_2$ (nm) & $L_3$ (nm) & $t_3$ (nm) & $P$ (nm) & $\eta_{11\times 11{\rm PW}}$ & $\eta_{21\times 21{\rm PW}}$ \cr
  \hline
  \hline
   $N=1$ stack & {314} & {124} & - & - & - & - & {419} & {88.8\%} & {88.7\%} \cr
  \hline
  \hline
   $N=2$ stacks & {192} & {121} & {332} & {127} & - & - & {369} & {98.5\%} & {98.5\%} \cr
  \hline
  \hline
   $N=3$ stacks & {151} & {120} & {277} & {127} & {427} & {123} & {467} & {99.4\%} & {\bf 99.1\%} \cr
  \hline
  \end{tabular}
 \end{center}
\begin{center}
  Ti / PMMA\vskip 7pt
  \begin{tabular}{|l|c|c|c|c|c|c|c||c|c|}
  \hline
     & $L_1$ (nm) & $t_1$ (nm) & $L_2$ (nm) & $t_2$ (nm) & $L_3$ (nm) & $t_3$ (nm) & $P$ (nm) & $\eta_{11\times 11{\rm PW}}$ & $\eta_{21\times 21{\rm PW}}$ \cr
  \hline
  \hline
   $N=1$ stack & {368} & {134} & - & - & - & - & {491} & {89.3\%} & {88.9\%} \cr
  \hline
  \hline
   $N=2$ stacks & {175} & {117} & {306} & {127} & - & - & {336} & {98.6\%} & {98.5\%} \cr
  \hline
  \hline
   $N=3$ stacks & {161} & {125} & {302} & {126} & {451} & {113} & {491} & {99.4\%} & {\bf 99.0\%} \cr
   \hline
  \end{tabular}
 \end{center}
 }
 \caption{\label{table1}Geometrical parameters and figure of merit $\eta$ for optimal structures made of $N=1$, 2 or 3 stacks of Ni/PMMA, W/PMMA, Cr/PMMA or Ti/PMMA (from top to bottom). PW: number of plane waves used in the calculations of $\eta$.}
\end{table}


\begin{table}[t]
 \vspace{0.4cm}
 {\scriptsize
 \begin{center}
  Al / PMMA\vskip 7pt
  \begin{tabular}{|l|c|c|c|c|c|c|c||c|c|}
  \hline
     & $L_1$ (nm) & $t_1$ (nm) & $L_2$ (nm) & $t_2$ (nm) & $L_3$ (nm) & $t_3$ (nm) & $P$ (nm) & $\eta_{11\times 11{\rm PW}}$ & $\eta_{21\times 21{\rm PW}}$ \cr
  \hline
  \hline
   $N=1$ stack & {221} & {151} & - & - & - & - & {491} & {89.2\%} & {\bf 63.8\%} \cr
  \hline
  \hline
   $N=2$ stacks & {187} & {108} & {263} & {58} & - & - & {415} & {97.9\%} & {\bf 81.0\%} \cr
  \hline
  \hline
   $N=3$ stacks & {129} & {112} & {208} & {110} & {294} & {97} & {465} & {98.7\%} & {\bf 83.6\%} \cr
  \hline
  \end{tabular}
 \end{center}
 \begin{center}
  Cu / PMMA\vskip 7pt
  \begin{tabular}{|l|c|c|c|c|c|c|c||c|c|}
  \hline
     & $L_1$ (nm) & $t_1$ (nm) & $L_2$ (nm) & $t_2$ (nm) & $L_3$ (nm) & $t_3$ (nm) & $P$ (nm) & $\eta_{11\times 11{\rm PW}}$ & $\eta_{21\times 21{\rm PW}}$ \cr
  \hline
  \hline
   $N=1$ stack & {149} & {112} & - & - & - & - & {230} & {80.9\%} & {\bf 55.2\%} \cr
  \hline
  \hline
   $N=2$ stacks & {188} & {127} & {283} & {120} & - & - & {447} & {89.1\%} & {\bf 75.8\%} \cr
  \hline
  \hline
   $N=3$ stacks & {119} & {145} & {211} & {132} & {299} & {111} & {365} & {95.5\%} & {\bf 77.5\%} \cr
  \hline
  \end{tabular}
 \end{center}
 \begin{center}
  Au / PMMA\vskip 7pt
  \begin{tabular}{|l|c|c|c|c|c|c|c||c|c|}
  \hline
     & $L_1$ (nm) & $t_1$ (nm) & $L_2$ (nm) & $t_2$ (nm) & $L_3$ (nm) & $t_3$ (nm) & $P$ (nm) & $\eta_{11\times 11{\rm PW}}$ & $\eta_{21\times 21{\rm PW}}$ \cr
  \hline
  \hline
   $N=1$ stack & {196} & {160} & - & - & - & - & {416} & {68.5\%} & {\bf 52.5\%} \cr
  \hline
  \hline
   $N=2$ stacks & {158} & {152} & {267} & {89} & - & - & {326} & {85.5\%} & {\bf 65.1\%} \cr
  \hline
  \hline
   $N=3$ stacks & {172} & {79} & {202} & {110} & {335} & {100} & {409} & {91.2\%} & {\bf 70.4\%} \cr
  \hline
  \end{tabular}
 \end{center}
 \begin{center}
  Ag / PMMA\vskip 7pt
  \begin{tabular}{|l|c|c|c|c|c|c|c||c|c|}
  \hline
     & $L_1$ (nm) & $t_1$ (nm) & $L_2$ (nm) & $t_2$ (nm) & $L_3$ (nm) & $t_3$ (nm) & $P$ (nm) & $\eta_{11\times 11{\rm PW}}$ & $\eta_{21\times 21{\rm PW}}$ \cr
  \hline
  \hline
   $N=1$ stack & {152} & {112} & - & - & - & - & {235} & {43.9\%} & {\bf 35.8\%} \cr
  \hline
  \hline
   $N=2$ stacks & {179} & {144} & {303} & {108} & - & - & {369} & {57.1\%} & {\bf 42.9\%} \cr
  \hline
  \hline
   $N=3$ stacks & {101} & {151} & {193} & {155} & {268} & {110} & {330} & {66.1\%} & {\bf 46.4\%} \cr
  \hline
  \end{tabular}
 \end{center}
 }
 \caption{\label{table2}Geometrical parameters and figure of merit $\eta$ for optimal structures made of $N=1$, 2 or 3 stacks of Al/PMMA, Cu/PMMA, Au/PMMA or Ag/PMMA (from top to bottom). PW: number of plane waves used in the calculations of $\eta$.}
\end{table}

\begin{figure}[t]
 \begin{center}
  \begin{tabular}{cc}
   \includegraphics[width=7.5cm]{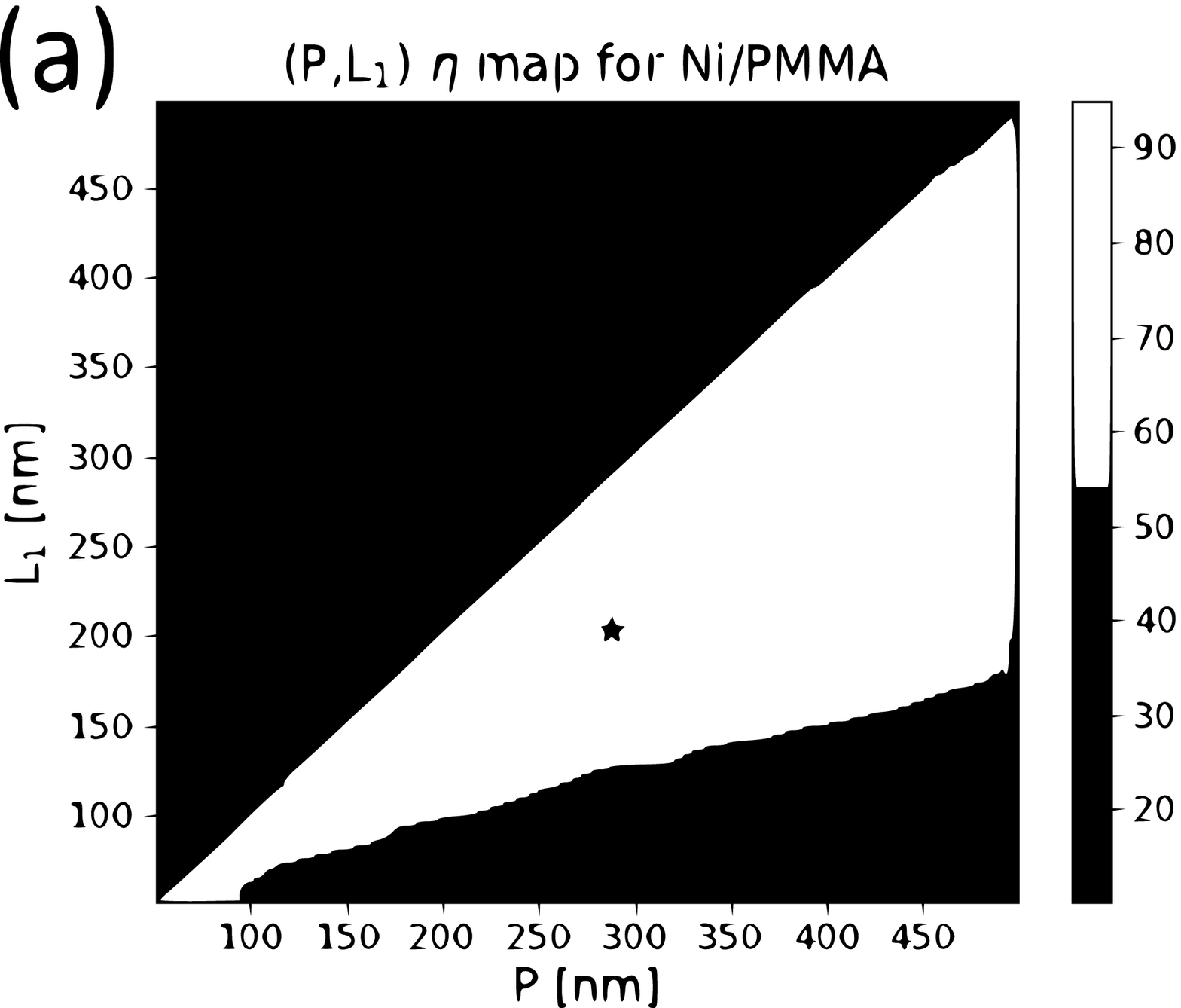} & \includegraphics[width=7.5cm]{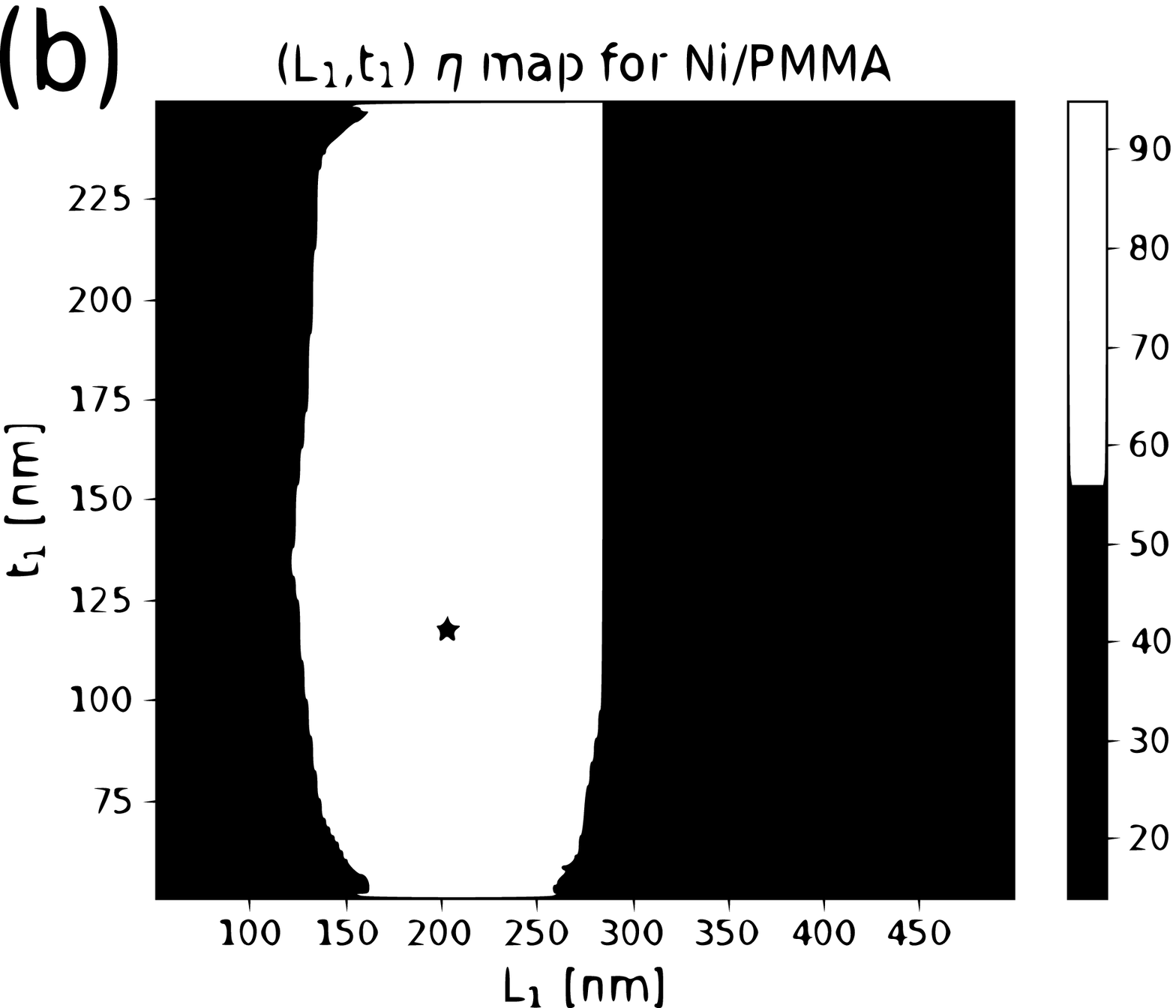}
  \end{tabular}
  \begin{tabular}{cc}
   \includegraphics[width=7.5cm]{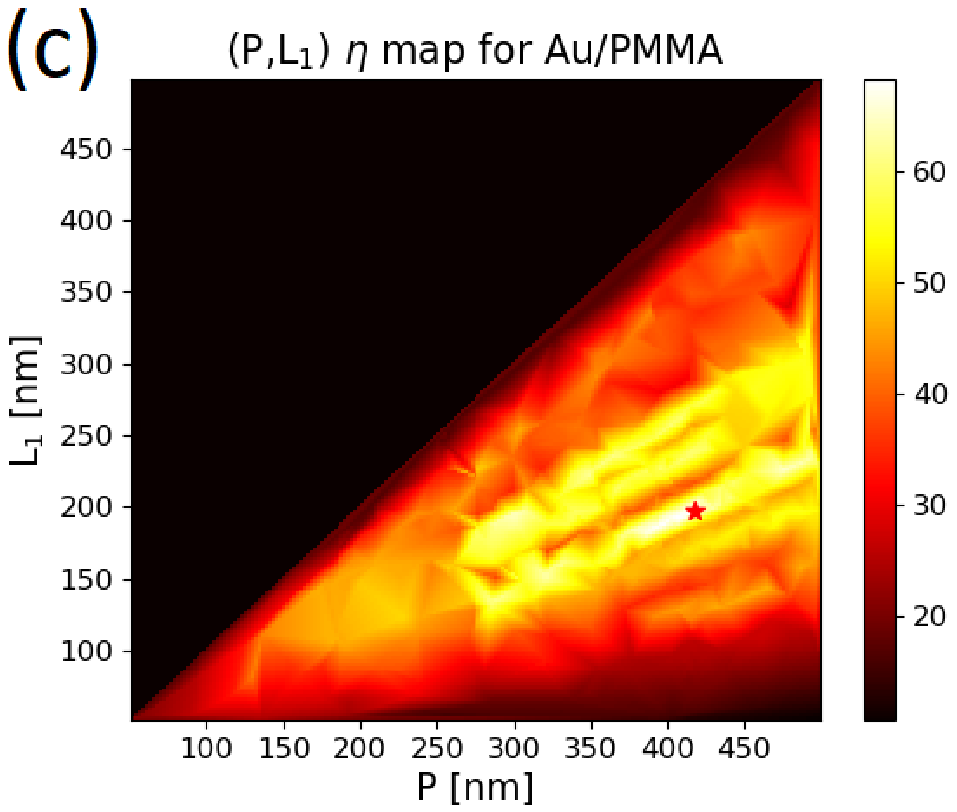} & \includegraphics[width=7.5cm]{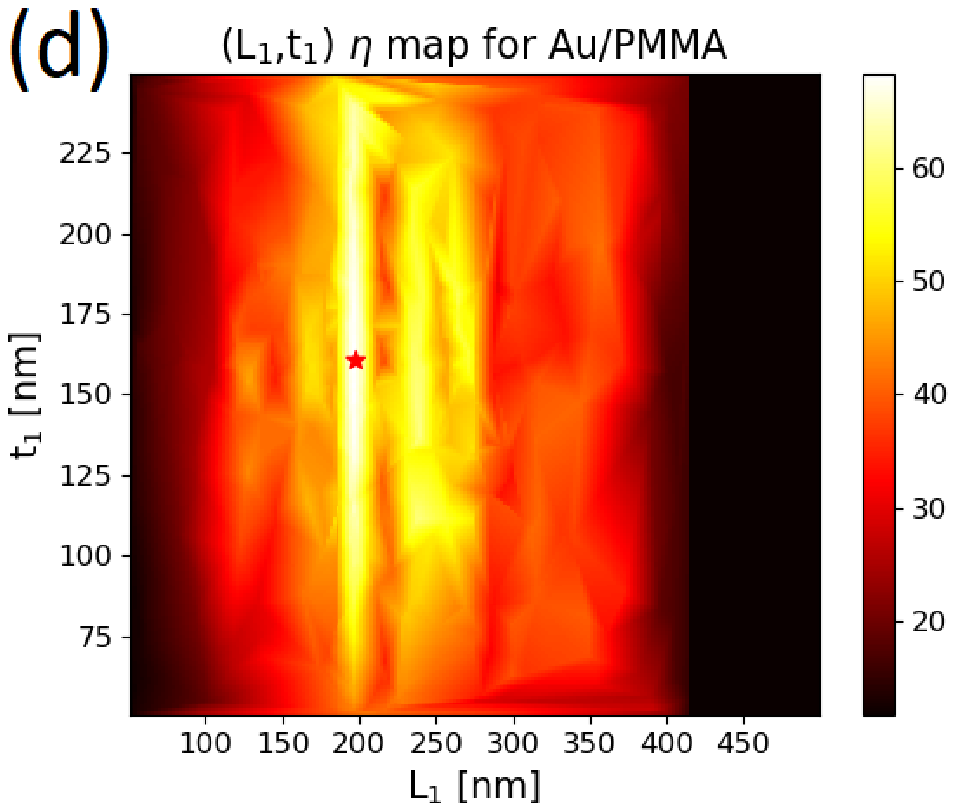}
  \end{tabular}
 \end{center}
 \caption{\label{figure5}Maps obtained by a linear interpolation of the data collected by
 the genetic algorithm in the $(P,L_1)$ and $(L_1,t_1)$ planes that include the
 final solution established by the algorithm (indicated by a star).
 (a)-(b) optimization of 1 stack of Ni/PMMA (global optimum at $P$=286 nm, $L_1$=201 nm and $t_1$=117 nm).
 (c)-(d) optimization of 1 stack of Au/PMMA (global optimum at $P$=416 nm, $L_1$=196 nm and $t_1$=160 nm).}
\end{figure}

High-quality solutions must actually meet two requirements in order to be easily implemented in real devices: (i) to provide the 
highest possible integrated absorptance $\eta$, but also (ii) to be robust with respect to
slight variations of the geometrical parameters (compatibility with fabrication tolerances).
We ideally want to find a broad optimum rather than a sharp one.
When running the GA to find optimal geometrical parameters, it is possible to interpolate the data collected by the algorithm and establish 
maps of the fitness in 2-D planes that cross the $n$-dimension point finally established by the GA. These maps reveal the robustness of the final solution 
(i.e., its stability with respect to slight variations of the geometrical parameters).
An example is given in Fig. \ref{figure5}, where we compare the maps obtained when optimizing $N=1$ stack of Ni/PMMA (broad optimum - Fig. \ref{figure5}(a)-(b-) and $N=1$ stack of Au/PMMA
(sharp optimum - Fig. \ref{figure5}(c)-(d)). These maps reveal that the solution found for Ni/PMMA is actually robust for practical applications, while the solution found for 
Au/PMMA is too sensitive. For $N$=1 stack of Ni/PMMA, $P$, $L_1$ and $t_1$ have indeed a domain of variation of 82 nm, 51 nm and 71 nm respectively
around the optimum (domain associated with $\eta\geq \eta_{\rm opt}-5\%$). For $N$=1 stack of Au/PMMA, this domain is reduced 
to 21 nm, 10 nm and 91 nm respectively. The lateral dimension $L_1$ is the most sensitive parameter in this case.
All solutions presented in Table \ref{table1} were checked for their robustness, based on the inspection of the fitness 
maps established by the GA. They appeared to correspond to broad optima suitable for practical applications. On the contrary, the solutions 
presented in Table \ref{table2} were discarded because they actually correspond to sharp optima.

A quality check of the reliability of the presented solutions is performed by increasing the plane wave number to $21\times21$ (see Table 1 and 2 and extended discussion in SM). This quality check allows us to further confirm the stability of our results: while increasing the plane wave number, good solutions (Table 1 - Ni;W;Cr;Ti) are stable regarding the $\eta$ parameter (deviation of $\eta$ limited to 1.2\%, when comparing 11$\times$11 and 21$\times$21 PW), while less optimal ones (Table 2 - Al;Cu;Au;Ag) are much less stable (deviation of $\eta$ between 8.1\% and 25.7\%, when comparing 11$\times$11 and 21$\times$21 PW). 

\begin{figure}[t]
 \begin{center}
   \includegraphics[width=10cm]{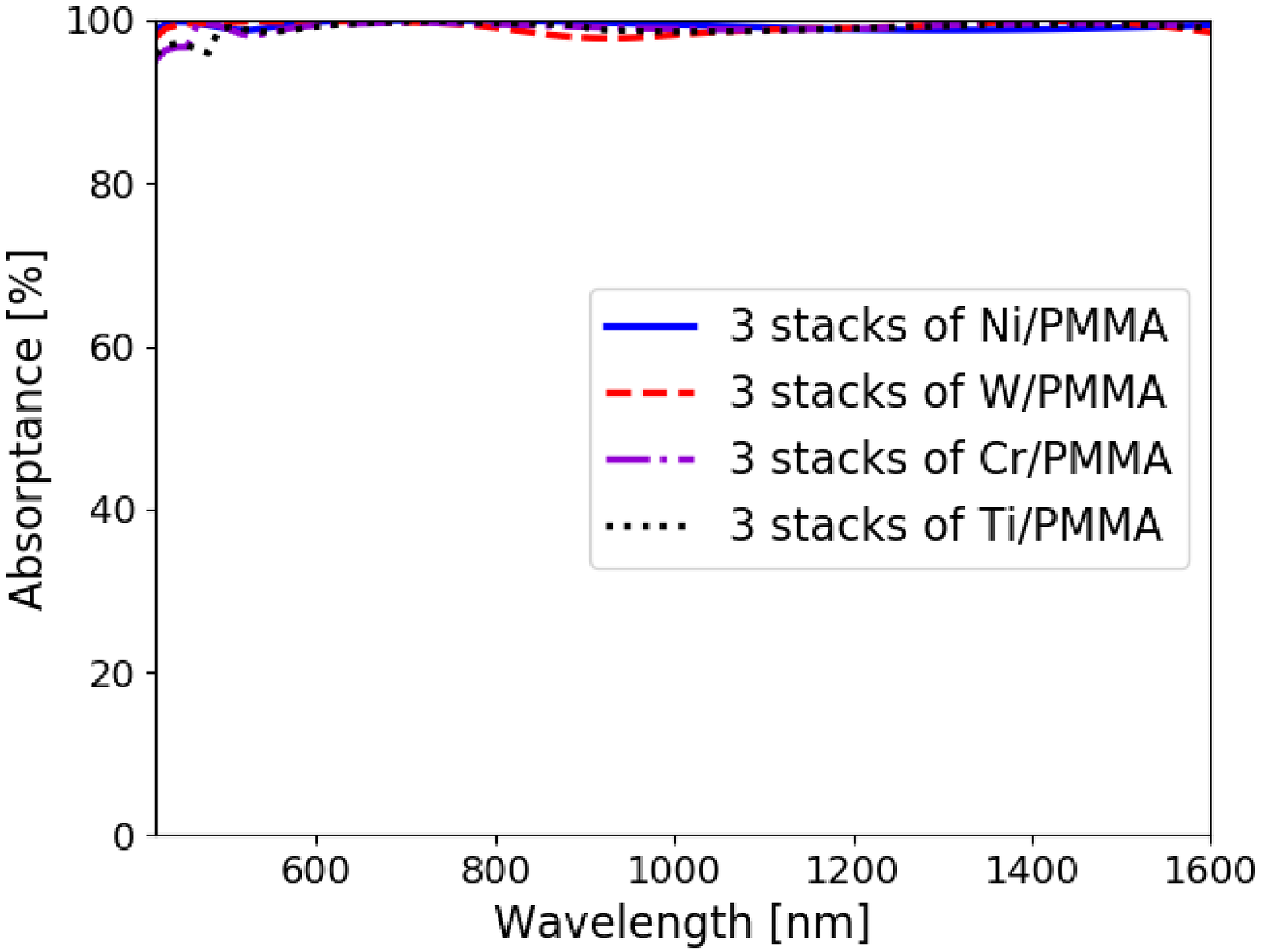}
 \end{center}
 \caption{\label{figure6}Absorptance spectrum for optimized structures made of $N=3$ stacks of Ni/PMMA (solid), W/PMMA (dashed), 
   Cr/PMMA (dot-dashed) and Ti/PMMA (dotted). These results correspond to a normally incident radiation and RCWA calculations 
   using $21 \times 21$ plane waves.}
\end{figure}

Fig. \ref{figure6} represents the absorptance spectrum obtained for normally incident radiations with the four best structures identified 
in this work ($N=3$ stacks of Ni/PMMA, W/PMMA, Cr/PMMA or Ti/PMMA). These structures provide respectively an integrated absorptance $\eta$
of 99.3\%, 99.2\%, 99.1\% and 99.0\% (results for RCWA calculations using 21$\times$21 plane waves).
These results are amongst the best reported so far in the literature and confirm that the initial goal of designing an ultra-broadband PEA in the visible-NIR range is met.
The challenge of fabricating such PEA is mitigated by the small number of layers and the selection of metals that are better alternatives to noble metals. Moreover, the proposed PEA are robust with respect to fabrication tolerances.

\subsection{Plasmonic absorber characteristics}

In order to better describe the physics of those PEA, we will focus from now on only on the best structure identified in this work, i.e. the one made of $N=3$ stacks of Ni/PMMA. It provides an integrated absorptance of 99.3\%. As shown above, the solution is robust with respect to deviations of the geometrical parameters
(broad optimum), as confirmed by 2-D maps of the fitness around the optimum and the comparison between the results obtained with 
11$\times$11 and 21$\times$21 plane waves.

We can calculate the Poynting vector 
$\vec S = \frac{1}{2}\ \vec E\times \vec H^*$, where $\vec E$ and $\vec H$ refer here to the complex-number representation of the electric and magnetic fields, in order to show the energy flow through the structure. Moreover, we can determine the local absorption inside the PEA.
Based on the method developed by Brenner,\cite{Brenner_2010} the absorbed power $P_a$ in a volume $V$ can be estimated by
\begin{equation}
P_a=\frac{\varepsilon_0\omega}{2}\int_{V} Im(\varepsilon(\bm{r},\omega))|\bm{E(\bm{r},\omega)}|^2 dV
\label{P_a}
\end{equation}
with $\varepsilon(\bm{r},\omega)$ the local complex electric permittivity. By normalizing the absorbed power $P_a$ to the incident power $P_i$, one obtains the local absorptance $A_{loc}(\bm{r},\omega)=P_a/P_i$.

\begin{figure}[t]
 \begin{center}
  \begin{tabular}{cc}
   \includegraphics[width=8.0cm]{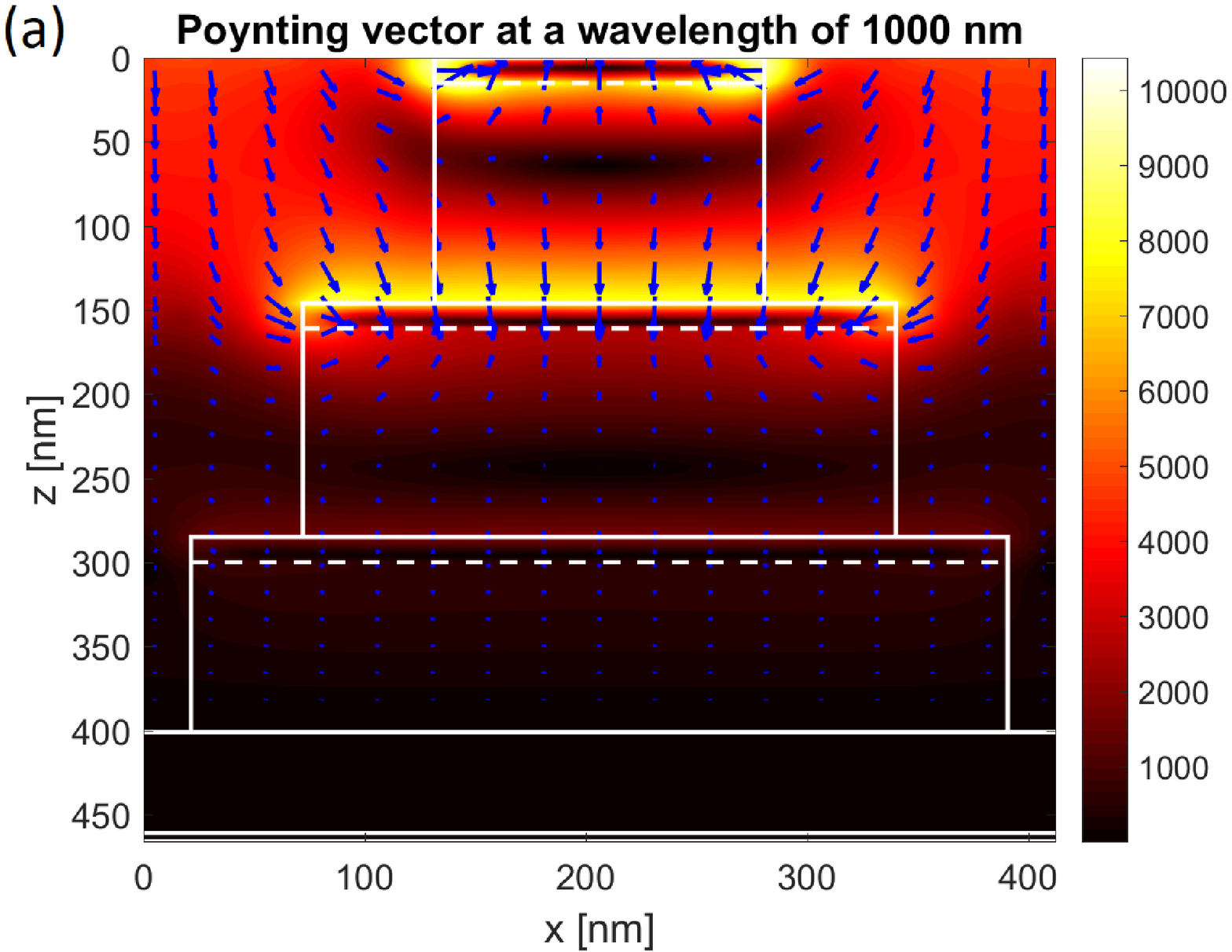} & \includegraphics[width=8.0cm]{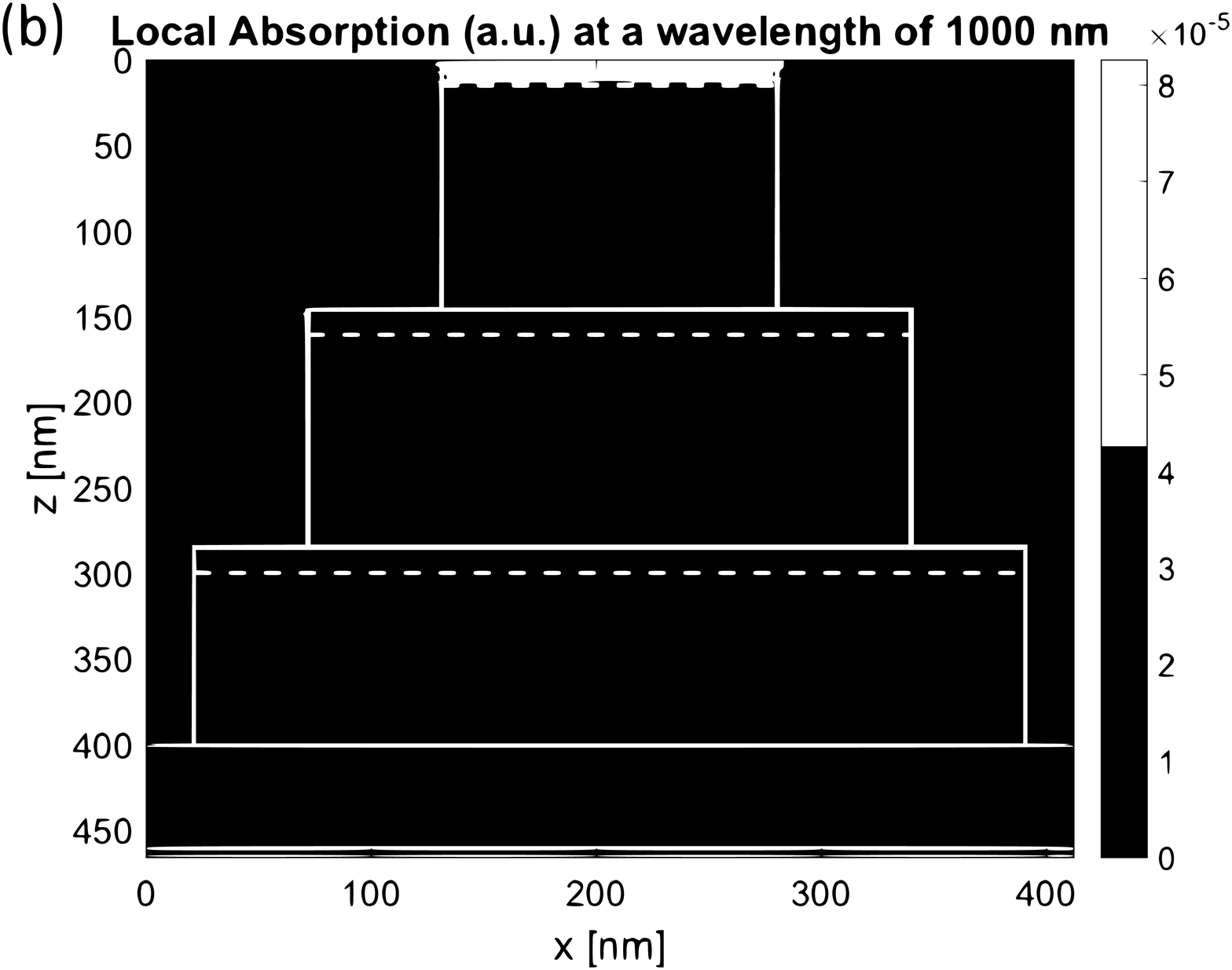}
  \end{tabular}
 \end{center}
 \caption{\label{figure7}(a) Time-averaged Poynting vector and (b) time-averaged local absorption 
 in the central $(x,z)$ plane of a structure made of $N=3$ stacks of Ni/PMMA for a normally incident radiation at
 a wavelength of 1000 nm.}
\end{figure}

These calculations provide additional physical insight by showing in which parts of the structure the incident radiations are absorbed.
An example is provided in Fig.\ref{figure7} for a normally incident radiation at a wavelength of 1000 nm. These results show that 
this radiation is essentially absorbed in the second metallic layer (25.9\% of the incident radiation is absorbed in the 
top metallic layer, 50.3\% in the central metallic layer and 23.3\% in the bottom metallic layer).
This is totally consistent with our previous work showing that the lower (upper) part of the pyramid absorbs higher (lower) wavelength \cite{Lobet_2014}, as predicted by the variation of LSP with the size of the resonators.

\begin{figure}[t]
 \begin{center}
  \begin{tabular}{cc}
   \includegraphics[width=6.5cm]{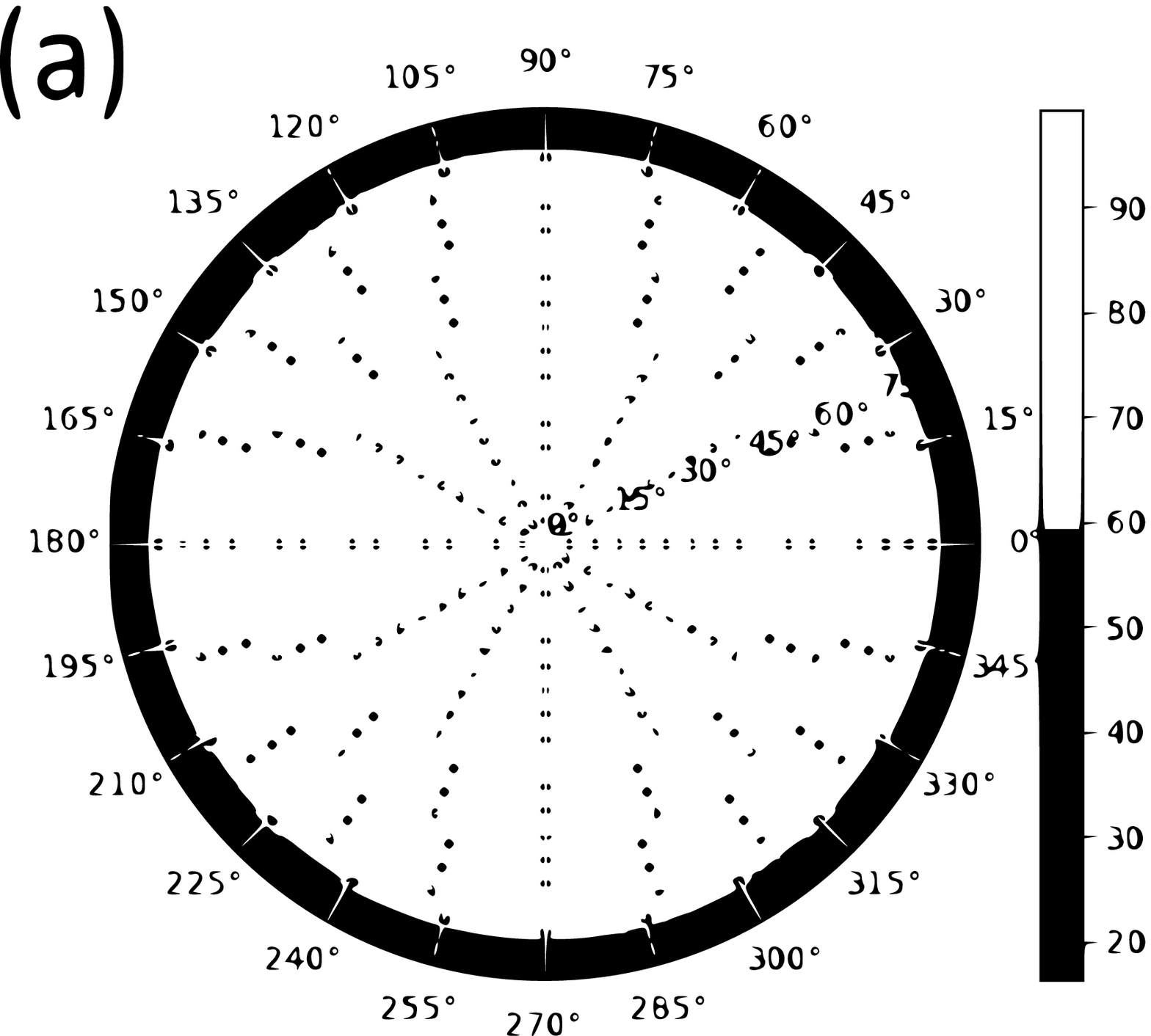} & \includegraphics[width=7.5cm]{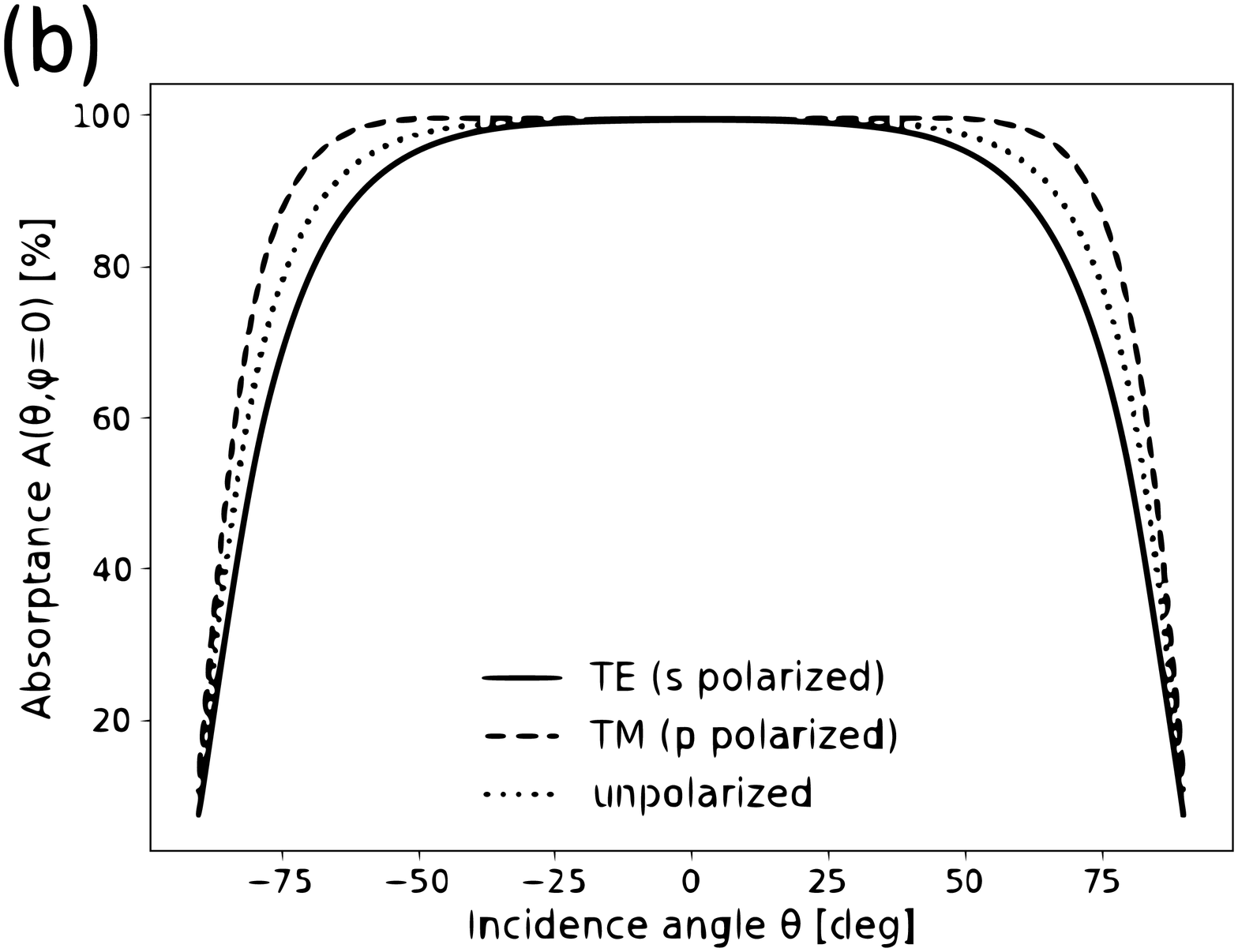}
  \end{tabular}
 \end{center}
 \caption{\label{figure8}Integrated absorptance for $N$=3 stacks of Ni/PMMA (results obtained with $21\times 21$ plane waves). Left: integrated absorptance for an unpolarized incident radiation, as a function of the polar and azimuthal angles $\theta$ and $\phi$. The blue points show the $(\theta,\phi)$ values 
 for which an explicit RCWA calculation was performed; these points are used in a cubic interpolation to generate a 2-D map.
 Right: integrated absorptance for a TE (s polarized), a TM (p polarized) and an unpolarized incident radiation, as a function of $\theta$ with $\phi=0$ (horizontal profile).}
\end{figure}

\begin{figure}[t]
 \begin{center}
  \includegraphics[width=10cm]{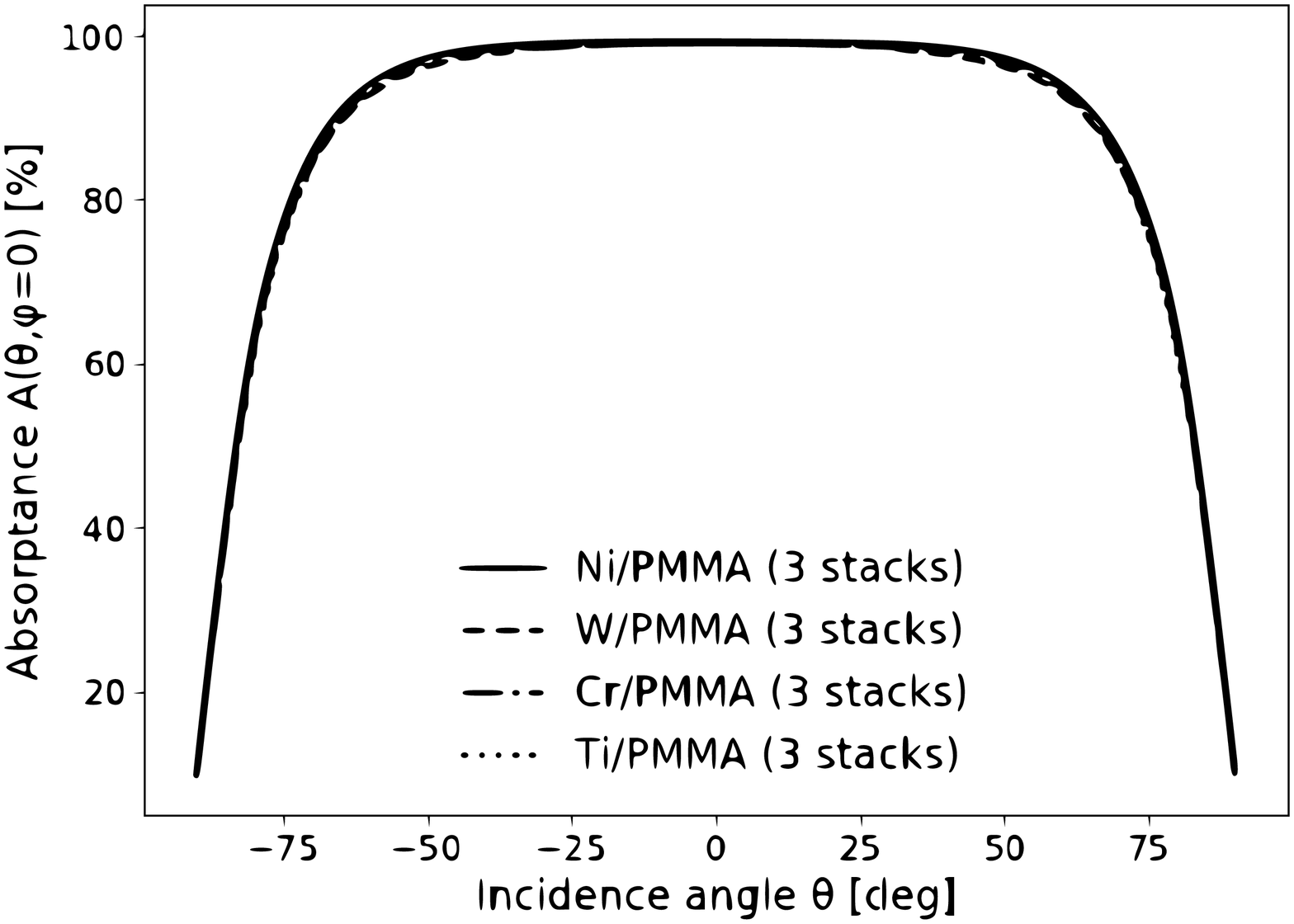}
 \end{center}
 \caption{\label{figure9}Integrated absorptance for an unpolarized incident radiation with polar angle $\theta \in [-90^\circ,90^\circ]$
 and azimuthal angle $\phi=0^\circ$ (horizontal profile). The results correspond to $N$=3 stacks of Ni/PMMA (solid), W/PMMA (dashed), Cr/PMMA (dot-dashed)
 and Ti/PMMA (dotted). These results are obtained with 21$\times$21 plane waves.}
\end{figure}

The results presented so far concerned the absorption of normally incident radiations. However, in many applications such as stray light mitigation, the PEA should be stable from an angular point of view. 
Therefore, we check how the integrated absorptance $\eta$
for $N=3$ stacks of Ni/PMMA depends on the polar and azimuthal angles ($\theta$ and $\phi$) of the incident radiation.
These results are presented in Fig. \ref{figure8}. The left part of this figure provides a complete 2-D map of the integrated absorptance
with respect to the polar and azimuthal angles of the incident radiation. This map is actually generated by a cubic interpolation of data computed
for only 73 points (blue points in the representation). The right part of this figure shows an horizontal section 
($\theta \in [-90^\circ,90^\circ]$ and $\phi=0^\circ$) for incident radiations that are either $s-$polarized (TE), $p-$polarized (TM) or
unpolarized ($\frac 1 2$(TE+TM)). These results show that the integrated absorptance $\eta(\theta,\phi)$
remains quasi-perfect over a broad angular domain
($\eta(\theta,\phi)\geq 98.5\%$ for $\theta\leq 40^\circ$). This makes the proposed solution for 
a broadband quasi-perfect absorber made of three stacks of Ni/PMMA extremely robust for practical applications also from an angular point of view.
Fig. \ref{figure9} finally shows the horizontal profiles achieved when considering $N=3$ stacks 
of Ni/PMMA, W/PMMA, Cr/PMMA or Ti/PMMA, i.e. for the four best solutions identified
in this work, in the case of an unpolarized incident radiation. This figure confirms the exceptional 
robustness of these best four solutions with respect to the inclination of the incident radiation.

\section{Conclusion}

Numerical investigations of truncated square-based pyramids made of a few number (one to three) of
alternating stacks of metal/dielectric layers are carried out. We focused on realistic configurations consisting of maximum three metallic layers stacked above each other. The GA strategy allowed to explore $10^{17}$ geometrical configurations and selected the optimal ones. Ni/PMMA, W/PMMA, Cr/PMMA and Ti/PMMA are determined as the best configurations for realizing PEA over the visible and NIR range, with integrated absorption higher than 99 \% once three layers are considered. Those PEA are robust against geometrical parameter deviations that might occur during experimental realization. 
Moreover, these optimal solutions maintain quasi-perfect broadband absorption properties over a broad angular range when changing the incidence angle of EM radiation. This study offers guidelines for a realistic design of PEA, that can readily be fabricated using currently available micro/nanofabrication techniques, using modest resources. 

\begin{acknowledgement}
The authors would like to thank N. Reckinger and L. Henrard for stimulating discussions. A.M. and B.H. are funded by the Fund for Scientific Research (F.R.S.-FNRS) of
Belgium. A.M. is member of NaXys, Namur Institute for Complex Systems,
University of Namur, Belgium.
This research used resources of the ``Plateforme Technologique de Calcul Intensif (PTCI)''
(http://www.ptci.unamur.be) located at the University of Namur, Belgium, which is supported by 
the F.R.S.-FNRS under the convention No. 2.5020.11. The PTCI is member of the ``Consortium des
Equipements de Calcul Intensif (CECI)'' (http://www.ceci-hpc.be).
The present research also benefited from computational resources made available on the
Tier-1 supercomputer of the F\'ed\'eration Wallonie-Bruxelles, infrastructure funded by the
Walloon Region under the grant agreement No. 1117545. Part of this work was performed while M.L. was a recipient of a Fellowship of the Belgian American Educational Foundation.
\end{acknowledgement}

\begin{suppinfo}

{\bf Range of variation of the parameters for the Genetic Algorithm}

We consider that the lateral periodicity $P$ of the system can
take values between 50 and 500 nm (by steps of 1 nm). The lateral dimension $L_i$ of each stack of metal/dielectric
layers can take values between 50 and 500 nm (by steps of 1 nm). The thickness $t_i$ of each dielectric
can take values between 50 and 250 nm (by steps of 1 nm). The subscripts $i$=1, 2 and 3 refer respectively to the 
stack at the apex, in the middle or at the bottom of each nanopyramid (Fig. 1). In order to obtain pyramidal structures, 
we require that the final solution satisfies
$L_1 < L_2 < L_3 \leq P-40{\rm\ nm}$, where 40 nm represents the minimal imposed safe distance 
between adjacent pyramids for insuring realistic fabrication.
When optimizing structures made of three stacks of metal/dielectric layers, there are actually seven parameters to determine
($P$, $L_1$, $t_1$, $L_2$, $t_2$, $L_3$ and $t_3$), with a total of 13,936,405,106,594,025 possible parameter combinations
to consider if the relaxed constraint $L_1 < L_2 < L_3 \leq P$ is actually enforced during the optimization.

{\bf Description of the Genetic Algorithm}

Given $n$ decision variables $x_i \in [x_i^{\rm min}, x_i^{\rm max}]$ to determine (within a precision $\Delta x_i$
representative of experimental constrains on the fabrication of a device), the objective is to find the global
maximum of an objective function $f=f(x_1,\ldots x_n)$.
The variables $x_i$ are encoded by sequences of binary digits (genes), which actually represent in the original Gray code
the number of steps
$(x_i-x_i^{\rm min})/\Delta x_i$ between $x_i^{\rm min}$ and $x_i$.\cite{Eiben_2007}
We refer by DNA to a complete set of $n$ genes.
We work with a population of $n_{\rm pop}$=50 individuals. The initial population consists of random individuals.
At each generation, we evaluate in parallel the fitness $f(x_1,\ldots x_n)$ of new individuals. We keep a record with
all fitness evaluations in order to avoid any duplication of these evaluations. The population is sorted
from the best individual to the worst. The worst $n_{\rm rand}$ individuals are replaced by random individuals
in the next generation. We use $n_{\rm rand}=0.1\times n_{\rm pop} \times (1-p)$, where $p=|s-0.5|/0.5$ is a progress
indicator and $s$ is the genetic similarity (fraction of bits in the population whose value is identical to the best
individual). The remaining part of the population ($N=n_{\rm pop}-n_{\rm rand}$ individuals) participate to the steps of
selection, crossover and mutation.

The core operations of the Genetic Algorithm are the following.
{\it Selection:} $N$ parents are selected from a population of $N$ individuals by a rank-based roulette wheel selection,
noting that a given individual can be selected several times.\cite{Eiben_2007}
{\it Crossover:} For any pair of parents, we define two children for the next generation either (i) by a
crossover operation (probability of 70\%), or (ii) by a simple replication of the parents (probability of 30\%).
In the current version of our GA, the crossover operation can be a binary one-point crossover\footnote{In a binary
one-point crossover, the first $n_{\rm cut}$ bits of the DNA of the children come from one parent. The remaining
$n_{\rm bits}-n_{\rm cut}$ bits come from the other parent.
The point $n_{\rm cut}$ at which the parents' DNA is exchanged is chosen randomly in the interval $[1,n_{\rm bits}-1]$.} between the
DNA of the two parents\cite{Mayer_SPIE2020} (probability $p_{\rm bin}$ of 0.8 initially) or a real-valued
crossover\footnote{If ${\vec x}_1$ and ${\vec x}_2$ are the real variables represented by the two parents, the children
obtained by a real crossover between these parents will
represent a variable ${\vec x} = {\vec x}_1 + (2*{\rm rnd}-0.5)\times ({\vec x}_2-{\vec x}_1)$, where rnd is a
random number uniformly distributed in [0,1].}
between the variables ${\vec x}$ represented by the two parents (probability of $1-p_{\rm bin}$).
$p_{\rm bin}$ is adapted according to the success of these operators.
{\it Mutation:}
The children obtained by crossover are subjected to mutations. This operation consists of a random flipping of the binary digits
of a DNA. The probability of individual bit flips is set to $m=0.95/n_{\rm bits}$, where $n_{\rm bits}$ is the number of bits in a DNA.
In order to increase the diversity of the displacements generated by these mutations, we actually
express the gene values in randomly-shifted versions of the original Gray code and apply the mutations to these
encodings (see Appendix A of Ref. \citenum{Mayer_SPIE2018} for details). In the current version of our GA,
mutations can be "isotropic" (in this case, the mutation operator is applied $n$ times on a given DNA). The probability
$p_{\rm iso}$ to apply isotropic mutations is set to 0.2 initially. This value is adapted according to the success of this
operator.

In order to converge more rapidly to the final solution, we establish at each generation a quadratic approximation
of the fitness in the close neighborhood of the best-so-far individual (this approximation is based on the data
collected by the genetic algorithm). If the optimum of this approximation is within the specified boundaries,
it replaces the last random individual scheduled for the next generation (see Appendix B of Ref.
\citenum{Mayer_SPIE2018} for details).
The data collected by the algorithm is also used to establish 2-D maps of the fitness, by using dedicated
interpolation techniques. This is useful for monitoring the progress of the algorithm and for assessing the quality
of the final solution.

The fitness of all individuals scheduled for the next generation is finally computed in parallel. The new population
is sorted from the best individual to the worst. If the best individual of the new generation is
not as good as the best individual of the previous generation, the elite of that previous generation replaces
an individual chosen at random in the new generation. We repeat these different steps from generation to
generation until a termination criterion is met.

{\bf Quality check of the optimization results based on the plane wave number}

A final quality criterion is certainly the reliability of the presented results. In order to confirm the quality of our solutions, we increased the number
of plane waves in the RCWA calculations to 21$\times$21 (instead of 11$\times$11 when running the GA). The results obtained are given in Tables
\ref{table1} and \ref{table2}. 
The comparison between 11$\times$11 PW and 21$\times$21 PW in Table \ref{table1} reveals that the solutions selected on the basis of 
high $\eta$ values and high robustness are also stable with respect to this numerical test (only slight deviations between 
$\eta_{11\times 11{\rm PW}}$ and $\eta_{21\times 21{\rm PW}}$).
On the contrary, the solutions in Table \ref{table2} that were discarded, essentially because of the high sensitivity of $\eta$ with respect to
the geometrical parameters, turn out to be significantly affected by this increase of the number of plane waves used in the RCWA calculations
(large deviations between $\eta_{11\times 11{\rm PW}}$ and $\eta_{21\times 21{\rm PW}}$). It proves that the solutions given in Table \ref{table2}
were rightly discarded (they fail this last reliability criterion).
The fact that solutions that sit on sharp optima are also solutions that require a higher number of plane waves for an accurate calculation
is actually consistent. 
This observation suggests a simple criterion for testing the robustness of solutions (stability with respect to deviations of their 
geometrical parameters): testing the stability  with respect to the number of plane waves used for the calculation.
This approach does not require the calculation of 2-D maps. A single calculation based on an increased number of plane waves
may be sufficient to get a clue !

\end{suppinfo}

\bibliography{Library}

\providecommand{\latin}[1]{#1}
\makeatletter
\providecommand{\doi}
  {\begingroup\let\do\@makeother\dospecials
  \catcode`\{=1 \catcode`\}=2 \doi@aux}
\providecommand{\doi@aux}[1]{\endgroup\texttt{#1}}
\makeatother
\providecommand*\mcitethebibliography{\thebibliography}
\csname @ifundefined\endcsname{endmcitethebibliography}
  {\let\endmcitethebibliography\endthebibliography}{}
\begin{mcitethebibliography}{65}
\providecommand*\natexlab[1]{#1}
\providecommand*\mciteSetBstSublistMode[1]{}
\providecommand*\mciteSetBstMaxWidthForm[2]{}
\providecommand*\mciteBstWouldAddEndPuncttrue
  {\def\EndOfBibitem{\unskip.}}
\providecommand*\mciteBstWouldAddEndPunctfalse
  {\let\EndOfBibitem\relax}
\providecommand*\mciteSetBstMidEndSepPunct[3]{}
\providecommand*\mciteSetBstSublistLabelBeginEnd[3]{}
\providecommand*\EndOfBibitem{}
\mciteSetBstSublistMode{f}
\mciteSetBstMaxWidthForm{subitem}{(\alph{mcitesubitemcount})}
\mciteSetBstSublistLabelBeginEnd
  {\mcitemaxwidthsubitemform\space}
  {\relax}
  {\relax}

\bibitem[Planck(1914)]{Planck_1914}
Planck,~M. \emph{The theory of heat radiation}; P. Blakiston's Son \& Co:
  Philadelphia, 1914\relax
\mciteBstWouldAddEndPuncttrue
\mciteSetBstMidEndSepPunct{\mcitedefaultmidpunct}
{\mcitedefaultendpunct}{\mcitedefaultseppunct}\relax
\EndOfBibitem
\bibitem[Kirchhoff(1860)]{black_body}
Kirchhoff,~G. I. On the relation between the radiating and absorbing powers of
  different bodies for light and heat. \emph{The London, Edinburgh, and Dublin
  Philosophical Magazine and Journal of Science} \textbf{1860}, \emph{20},
  1--21\relax
\mciteBstWouldAddEndPuncttrue
\mciteSetBstMidEndSepPunct{\mcitedefaultmidpunct}
{\mcitedefaultendpunct}{\mcitedefaultseppunct}\relax
\EndOfBibitem
\bibitem[Lummer and Kurlbaum(1898)Lummer, and Kurlbaum]{Lummer_1898}
Lummer,~O.; Kurlbaum,~F. Der electrisch geglühte "absolut schwarze" Körper
  und seine Temperaturmessung. \emph{Verhandlungen der Deutschen Physikalischen
  Gesellschaft} \textbf{1898}, \emph{17}, 106--111\relax
\mciteBstWouldAddEndPuncttrue
\mciteSetBstMidEndSepPunct{\mcitedefaultmidpunct}
{\mcitedefaultendpunct}{\mcitedefaultseppunct}\relax
\EndOfBibitem
\bibitem[Lummer and Kurlbaum(1901)Lummer, and Kurlbaum]{Lummer_1901}
Lummer,~O.; Kurlbaum,~F. Der elektrisch geglühte "schwarze" Körper.
  \emph{Annalen der Physik} \textbf{1901}, \emph{310}, 829--836\relax
\mciteBstWouldAddEndPuncttrue
\mciteSetBstMidEndSepPunct{\mcitedefaultmidpunct}
{\mcitedefaultendpunct}{\mcitedefaultseppunct}\relax
\EndOfBibitem
\bibitem[Mehra and Rechenberg(2000)Mehra, and Rechenberg]{Mehra_2000}
Mehra,~J.; Rechenberg,~H. \emph{The historical development of quantum theory};
  Springer, 2000; p~39\relax
\mciteBstWouldAddEndPuncttrue
\mciteSetBstMidEndSepPunct{\mcitedefaultmidpunct}
{\mcitedefaultendpunct}{\mcitedefaultseppunct}\relax
\EndOfBibitem
\bibitem[Barends \latin{et~al.}(2011)Barends, Wenner, Lenander, Chen, Bialczak,
  Kelly, Lucero, O’Malley, Mariantoni, Sank, Wang, White, Yin, Zhao, Cleland,
  Martinis, and Baselmans]{Barends_2011}
Barends,~R. \latin{et~al.}  Minimizing quasiparticle generation from stray
  infrared light in superconducting quantum circuits. \emph{Applied Physics
  Letters} \textbf{2011}, \emph{99}, 113507\relax
\mciteBstWouldAddEndPuncttrue
\mciteSetBstMidEndSepPunct{\mcitedefaultmidpunct}
{\mcitedefaultendpunct}{\mcitedefaultseppunct}\relax
\EndOfBibitem
\bibitem[Córcoles \latin{et~al.}(2011)Córcoles, Chow, Gambetta, Rigetti,
  Rozen, Keefe, Rothwell, Ketchen, and Steffen]{Corcoles_2011}
Córcoles,~A.; Chow,~J.; Gambetta,~J.; Rigetti,~C.; Rozen,~J.; Keefe,~G.;
  Rothwell,~M.; Ketchen,~M.; Steffen,~M. Protecting superconducting qubits from
  radiation. \textbf{2011}, \emph{99}, 181906\relax
\mciteBstWouldAddEndPuncttrue
\mciteSetBstMidEndSepPunct{\mcitedefaultmidpunct}
{\mcitedefaultendpunct}{\mcitedefaultseppunct}\relax
\EndOfBibitem
\bibitem[Mizuno \latin{et~al.}(2009)Mizuno, Ishii, Kishida, Hayamizu, Yasuda,
  Futaba, Yumura, and Hata]{Mizuno_2009}
Mizuno,~K.; Ishii,~J.; Kishida,~H.; Hayamizu,~Y.; Yasuda,~S.; Futaba,~D.;
  Yumura,~M.; Hata,~K. A black body absorber from vertically aligned
  single-walled carbon nanotubes. \emph{Proceedings of the National Academy of
  Sciences} \textbf{2009}, \emph{106}, 6044--6047\relax
\mciteBstWouldAddEndPuncttrue
\mciteSetBstMidEndSepPunct{\mcitedefaultmidpunct}
{\mcitedefaultendpunct}{\mcitedefaultseppunct}\relax
\EndOfBibitem
\bibitem[Ishii and Ono(2003)Ishii, and Ono]{Ishii_2003}
Ishii,~J.; Ono,~A. A Fourier‐Transform Spectrometer for Accurate Thermometric
  Applications at Low Temperatures. \emph{AIP Conference Proceedings}
  \textbf{2003}, \emph{684}, 705--710\relax
\mciteBstWouldAddEndPuncttrue
\mciteSetBstMidEndSepPunct{\mcitedefaultmidpunct}
{\mcitedefaultendpunct}{\mcitedefaultseppunct}\relax
\EndOfBibitem
\bibitem[Maier(2007)]{Maier_2007}
Maier,~S. \emph{Plasmonics: Fundamentals and Applications}; Springer,
  2007\relax
\mciteBstWouldAddEndPuncttrue
\mciteSetBstMidEndSepPunct{\mcitedefaultmidpunct}
{\mcitedefaultendpunct}{\mcitedefaultseppunct}\relax
\EndOfBibitem
\bibitem[Engheta and Ziolkowski(2006)Engheta, and Ziolkowski]{Engheta2006}
Engheta,~N.; Ziolkowski,~R.~W. \emph{Metamaterials: physics and engineering
  explorations}; John Wiley \& Sons, 2006\relax
\mciteBstWouldAddEndPuncttrue
\mciteSetBstMidEndSepPunct{\mcitedefaultmidpunct}
{\mcitedefaultendpunct}{\mcitedefaultseppunct}\relax
\EndOfBibitem
\bibitem[Joannopoulos(2008)]{Joannopoulos2008}
Joannopoulos,~J. D. J.~D. \emph{{Photonic crystals : molding the flow of
  light}}; Princeton University Press, 2008; p 286\relax
\mciteBstWouldAddEndPuncttrue
\mciteSetBstMidEndSepPunct{\mcitedefaultmidpunct}
{\mcitedefaultendpunct}{\mcitedefaultseppunct}\relax
\EndOfBibitem
\bibitem[Lodahl \latin{et~al.}(2015)Lodahl, Mahmoodian, and Stobbe]{Lodahl2015}
Lodahl,~P.; Mahmoodian,~S.; Stobbe,~S. {Interfacing single photons and single
  quantum dots with photonic nanostructures}. \emph{Reviews of Modern Physics}
  \textbf{2015}, \emph{87}, 347--400\relax
\mciteBstWouldAddEndPuncttrue
\mciteSetBstMidEndSepPunct{\mcitedefaultmidpunct}
{\mcitedefaultendpunct}{\mcitedefaultseppunct}\relax
\EndOfBibitem
\bibitem[Pendry(2000)]{Pendry2000}
Pendry,~J. Negative refraction makes a perfect lens. \emph{Phys. Rev. Lett.}
  \textbf{2000}, \emph{85}, 3966--3969\relax
\mciteBstWouldAddEndPuncttrue
\mciteSetBstMidEndSepPunct{\mcitedefaultmidpunct}
{\mcitedefaultendpunct}{\mcitedefaultseppunct}\relax
\EndOfBibitem
\bibitem[Pendry and Smith(2006)Pendry, and Smith]{Pendry2006}
Pendry,~J.; Smith,~D. Controlling Electromagnetic Fields. \emph{Science}
  \textbf{2006}, \emph{312}, 1780--1782\relax
\mciteBstWouldAddEndPuncttrue
\mciteSetBstMidEndSepPunct{\mcitedefaultmidpunct}
{\mcitedefaultendpunct}{\mcitedefaultseppunct}\relax
\EndOfBibitem
\bibitem[Liberal and Engheta(2017)Liberal, and Engheta]{Liberal_2017}
Liberal,~I.; Engheta,~N. Near-zero refractive index photonics. \emph{Nature
  Photonics} \textbf{2017}, \emph{11}, 149--158\relax
\mciteBstWouldAddEndPuncttrue
\mciteSetBstMidEndSepPunct{\mcitedefaultmidpunct}
{\mcitedefaultendpunct}{\mcitedefaultseppunct}\relax
\EndOfBibitem
\bibitem[Khurgin(2015)]{Khurgin_2015}
Khurgin,~J. How to deal with the loss in plasmonics and metamaterials.
  \emph{Nature Nanotech} \textbf{2015}, \emph{10}, 2--6\relax
\mciteBstWouldAddEndPuncttrue
\mciteSetBstMidEndSepPunct{\mcitedefaultmidpunct}
{\mcitedefaultendpunct}{\mcitedefaultseppunct}\relax
\EndOfBibitem
\bibitem[Landy \latin{et~al.}(2008)Landy, Sajuyigbe, Mock, Smith, and
  Padilla]{Landy_2008}
Landy,~N.; Sajuyigbe,~S.; Mock,~J.; Smith,~D.; Padilla,~W. Perfect Metamaterial
  Absorber. \emph{Phys. Rev. Lett.} \textbf{2008}, \emph{100}, 207402\relax
\mciteBstWouldAddEndPuncttrue
\mciteSetBstMidEndSepPunct{\mcitedefaultmidpunct}
{\mcitedefaultendpunct}{\mcitedefaultseppunct}\relax
\EndOfBibitem
\bibitem[Hedayati \latin{et~al.}(2011)Hedayati, Javaherirahim, Mozooni,
  Abdelaziz, Tavassolizadeh, Chakravadhanula, Zaporojtchenko, Strunkus, Faupel,
  and Elbahri]{Hedayati_2011}
Hedayati,~M.; Javaherirahim,~M.; Mozooni,~B.; Abdelaziz,~R.;
  Tavassolizadeh,~A.; Chakravadhanula,~V.; Zaporojtchenko,~V.; Strunkus,~T.;
  Faupel,~F.; Elbahri,~M. Design of a perfect black absorber at visible
  frequencies using plasmonic metamaterials. \emph{Adv. Mater.} \textbf{2011},
  \emph{23}, 5410--5414\relax
\mciteBstWouldAddEndPuncttrue
\mciteSetBstMidEndSepPunct{\mcitedefaultmidpunct}
{\mcitedefaultendpunct}{\mcitedefaultseppunct}\relax
\EndOfBibitem
\bibitem[Hedayati \latin{et~al.}(2012)Hedayati, Faupel, and
  Elbahri]{Hedayati_2012}
Hedayati,~M.; Faupel,~F.; Elbahri,~M. Tunable broadband plasmonic perfect
  absorber at visible frequency. \emph{Appl. Phys. A-Mater.} \textbf{2012},
  \emph{109}, 769--773\relax
\mciteBstWouldAddEndPuncttrue
\mciteSetBstMidEndSepPunct{\mcitedefaultmidpunct}
{\mcitedefaultendpunct}{\mcitedefaultseppunct}\relax
\EndOfBibitem
\bibitem[Cui \latin{et~al.}(2011)Cui, Fung, Xu, Yi, He, and Fang]{Cui_2011}
Cui,~Y.; Fung,~K.; Xu,~J.; Yi,~J.; He,~S.; Fang,~N. Exciting multiple plasmonic
  resonances by a double-layered metallic nanostructure. \emph{J. Opt. Soc. Am.
  B} \textbf{2011}, \emph{28}, 2827--2832\relax
\mciteBstWouldAddEndPuncttrue
\mciteSetBstMidEndSepPunct{\mcitedefaultmidpunct}
{\mcitedefaultendpunct}{\mcitedefaultseppunct}\relax
\EndOfBibitem
\bibitem[Cui \latin{et~al.}(2012)Cui, Fung, Xu, He, and Fang]{Cui_2012}
Cui,~Y.; Fung,~K.; Xu,~J.; He,~S.; Fang,~N. Multiband plasmonic absorber based
  on transverse phase resonances. \emph{Opt. Express} \textbf{2012}, \emph{20},
  17552--17559\relax
\mciteBstWouldAddEndPuncttrue
\mciteSetBstMidEndSepPunct{\mcitedefaultmidpunct}
{\mcitedefaultendpunct}{\mcitedefaultseppunct}\relax
\EndOfBibitem
\bibitem[Zhu and Guo(2012)Zhu, and Guo]{Zhu_2012}
Zhu,~P.; Guo,~L. High performance broadband absorber in the visible band by
  engineered dispersion and geometry of a metal-dielectric-metal stack.
  \emph{Appl. Phys. Lett.} \textbf{2012}, \emph{101}, 241116\relax
\mciteBstWouldAddEndPuncttrue
\mciteSetBstMidEndSepPunct{\mcitedefaultmidpunct}
{\mcitedefaultendpunct}{\mcitedefaultseppunct}\relax
\EndOfBibitem
\bibitem[Wang \latin{et~al.}(2012)Wang, Sun, Paudel, Zhang, Ren, and
  Kempa]{Wang_2012}
Wang,~Y.; Sun,~T.; Paudel,~T.; Zhang,~Y.; Ren,~Z.; Kempa,~K.
  Metamaterial-plasmonic absorber structure for high efficiency amorphous
  silicon solar cells. \emph{Nano Lett.} \textbf{2012}, \emph{12},
  440--445\relax
\mciteBstWouldAddEndPuncttrue
\mciteSetBstMidEndSepPunct{\mcitedefaultmidpunct}
{\mcitedefaultendpunct}{\mcitedefaultseppunct}\relax
\EndOfBibitem
\bibitem[Ye \latin{et~al.}(2010)Ye, Jin, and He]{Ye_2010}
Ye,~Y.; Jin,~Y.; He,~S. Omnidirectional, polarization-insensitive and broadband
  thin absorber in the terahertz regime. \emph{J. Opt. Soc. Am. B}
  \textbf{2010}, \emph{27}, 498\relax
\mciteBstWouldAddEndPuncttrue
\mciteSetBstMidEndSepPunct{\mcitedefaultmidpunct}
{\mcitedefaultendpunct}{\mcitedefaultseppunct}\relax
\EndOfBibitem
\bibitem[Cui \latin{et~al.}(2012)Cui, Fung, Xu, Ma, Jin, He, and
  Fang]{Cui_2012b}
Cui,~Y.; Fung,~K.; Xu,~J.; Ma,~H.; Jin,~Y.; He,~S.; Fang,~N. Ultrabroadband
  light absorption by a sawtooth anisotropic metamaterial slab. \emph{Nano
  Lett.} \textbf{2012}, \emph{12}, 1443--1447\relax
\mciteBstWouldAddEndPuncttrue
\mciteSetBstMidEndSepPunct{\mcitedefaultmidpunct}
{\mcitedefaultendpunct}{\mcitedefaultseppunct}\relax
\EndOfBibitem
\bibitem[Ding \latin{et~al.}(2012)Ding, Cui, Ge, Jin, and He]{Ding_2012}
Ding,~F.; Cui,~Y.; Ge,~X.; Jin,~Y.; He,~S. Ultra-broadband microwave
  metamaterial absorber. \emph{Appl. Phys. Lett.} \textbf{2012}, \emph{100},
  103506\relax
\mciteBstWouldAddEndPuncttrue
\mciteSetBstMidEndSepPunct{\mcitedefaultmidpunct}
{\mcitedefaultendpunct}{\mcitedefaultseppunct}\relax
\EndOfBibitem
\bibitem[Argyropoulos \latin{et~al.}(2013)Argyropoulos, Le, Mattiucci,
  D'Aguanno, and Alu]{Argyropoulos_2013}
Argyropoulos,~C.; Le,~K.; Mattiucci,~N.; D'Aguanno,~G.; Alu,~A. Broadband
  absorbers and selective emitters based on plasmonic Brewster metasurfaces.
  \emph{Phys. Rev. B} \textbf{2013}, \emph{87}, 205112\relax
\mciteBstWouldAddEndPuncttrue
\mciteSetBstMidEndSepPunct{\mcitedefaultmidpunct}
{\mcitedefaultendpunct}{\mcitedefaultseppunct}\relax
\EndOfBibitem
\bibitem[Qiuqun \latin{et~al.}(2013)Qiuqun, Weixing, Wencai, Taisheng, Jingli,
  Hongsheng, and Shaohua]{Qiuqun_2013}
Qiuqun,~L.; Weixing,~Y.; Wencai,~Z.; Taisheng,~W.; Jingli,~Z.; Hongsheng,~Z.;
  Shaohua,~T. Numerical study of the meta-nanopyramid array as efficient solar
  energy absorber. \emph{Opt. Mater. Express} \textbf{2013}, \emph{3},
  1187--1196\relax
\mciteBstWouldAddEndPuncttrue
\mciteSetBstMidEndSepPunct{\mcitedefaultmidpunct}
{\mcitedefaultendpunct}{\mcitedefaultseppunct}\relax
\EndOfBibitem
\bibitem[Zhou \latin{et~al.}(2021)Zhou, Qin, Liang, Meng, Xu, Smith, and
  Liu]{Zhou_2021}
Zhou,~Y.; Qin,~Z.; Liang,~Z.; Meng,~D.; Xu,~H.; Smith,~D.~R.; Liu,~Y.
  Ultra-broadband metamaterial absorbers from long to very long infrared
  regime. \emph{Light: Science \& Applications} \textbf{2021}, \emph{10},
  138\relax
\mciteBstWouldAddEndPuncttrue
\mciteSetBstMidEndSepPunct{\mcitedefaultmidpunct}
{\mcitedefaultendpunct}{\mcitedefaultseppunct}\relax
\EndOfBibitem
\bibitem[Dixon \latin{et~al.}(2020)Dixon, Montazeri, Shayegannia, Barnard,
  Cabrini, Matsuura, Holman, and Kherani]{Dixon_2020}
Dixon,~K.; Montazeri,~A.~O.; Shayegannia,~M.; Barnard,~E.~S.; Cabrini,~S.;
  Matsuura,~N.; Holman,~H.-Y.; Kherani,~N.~P. Tunable rainbow light trapping in
  ultrathin resonator arrays. \emph{Light: Science \& Applications}
  \textbf{2020}, \emph{9}, 138\relax
\mciteBstWouldAddEndPuncttrue
\mciteSetBstMidEndSepPunct{\mcitedefaultmidpunct}
{\mcitedefaultendpunct}{\mcitedefaultseppunct}\relax
\EndOfBibitem
\bibitem[Hajian \latin{et~al.}(2019)Hajian, Ghobadi, Butun, and
  Ozbay]{Hajian19}
Hajian,~H.; Ghobadi,~A.; Butun,~B.; Ozbay,~E. Active metamaterial nearly
  perfect light absorbers: a review. \emph{J. Opt. Soc. Am. B} \textbf{2019},
  \emph{36}, F131--F143\relax
\mciteBstWouldAddEndPuncttrue
\mciteSetBstMidEndSepPunct{\mcitedefaultmidpunct}
{\mcitedefaultendpunct}{\mcitedefaultseppunct}\relax
\EndOfBibitem
\bibitem[Huang \latin{et~al.}(2016)Huang, Liu, Zhu, Masala, Alarousu, Han, and
  Fratalocchi]{Huang_2016}
Huang,~J.; Liu,~C.; Zhu,~Y.; Masala,~S.; Alarousu,~E.; Han,~Y.; Fratalocchi,~A.
  Harnessing structural darkness in the visible and infrared wavelengths for a
  new source of light. \emph{Nature Nanotechnology} \textbf{2016}, \emph{11},
  60--66\relax
\mciteBstWouldAddEndPuncttrue
\mciteSetBstMidEndSepPunct{\mcitedefaultmidpunct}
{\mcitedefaultendpunct}{\mcitedefaultseppunct}\relax
\EndOfBibitem
\bibitem[Ziegler \latin{et~al.}(2020)Ziegler, Dathe, Pollok, Langenhorst,
  Hübner, Wang, and Schaaf]{Ziegler2020}
Ziegler,~M.; Dathe,~A.; Pollok,~K.; Langenhorst,~F.; Hübner,~U.; Wang,~D.;
  Schaaf,~P. Metastable Atomic Layer Deposition: 3D Self-Assembly toward
  Ultradark Materials. \emph{ACS Nano} \textbf{2020}, \emph{14},
  15023--15031\relax
\mciteBstWouldAddEndPuncttrue
\mciteSetBstMidEndSepPunct{\mcitedefaultmidpunct}
{\mcitedefaultendpunct}{\mcitedefaultseppunct}\relax
\EndOfBibitem
\bibitem[Mayer and Lobet(2018)Mayer, and Lobet]{Mayer_SPIE2018}
Mayer,~A.; Lobet,~M. {UV} to near-infrared broadband pyramidal absorbers via a
  genetic algorithm optimization approach. \emph{Proc. SPIE} \textbf{2018},
  \emph{10671}, 1067127--1--11\relax
\mciteBstWouldAddEndPuncttrue
\mciteSetBstMidEndSepPunct{\mcitedefaultmidpunct}
{\mcitedefaultendpunct}{\mcitedefaultseppunct}\relax
\EndOfBibitem
\bibitem[Mayer \latin{et~al.}(2020)Mayer, Griesse-Nascimento, Bi, Mazur, and
  Lobet]{Mayer_SPIE2020}
Mayer,~A.; Griesse-Nascimento,~S.; Bi,~H.; Mazur,~E.; Lobet,~M. Optimization by
  a genetic algorithm of pyramidal structures made of one, two or three stacks
  of metal/dielectric layers for a quasi-perfect broadband absorption of UV to
  near-infrared radiations. \emph{Proc. SPIE} \textbf{2020}, \emph{11344},
  113441L--1--13\relax
\mciteBstWouldAddEndPuncttrue
\mciteSetBstMidEndSepPunct{\mcitedefaultmidpunct}
{\mcitedefaultendpunct}{\mcitedefaultseppunct}\relax
\EndOfBibitem
\bibitem[Liu \latin{et~al.}(2020)Liu, Maier, and Li]{Liu_2020}
Liu,~C.; Maier,~S.; Li,~G. Genetic-Algorithm-Aided Meta-Atom Multiplication for
  Improved Absorption and Coloration in Nanophotonics. \emph{ACS Photonics}
  \textbf{2020}, \emph{7}, 1716–--1722\relax
\mciteBstWouldAddEndPuncttrue
\mciteSetBstMidEndSepPunct{\mcitedefaultmidpunct}
{\mcitedefaultendpunct}{\mcitedefaultseppunct}\relax
\EndOfBibitem
\bibitem[Gong \latin{et~al.}(2021)Gong, Kim, Larsson, Methling, Alden,
  Kristensson, Brackmann, Eschrich, Jager, Kiefer, and Ehn]{Gong_2021}
Gong,~M.; Kim,~H.; Larsson,~J.; Methling,~T.; Alden,~M.; Kristensson,~E.;
  Brackmann,~C.; Eschrich,~T.; Jager,~M.; Kiefer,~W.; Ehn,~A. Fiber-based stray
  light suppression in spectroscopy using periodic shadowing. \emph{Opt.
  Express} \textbf{2021}, \emph{29}, 7232--7246\relax
\mciteBstWouldAddEndPuncttrue
\mciteSetBstMidEndSepPunct{\mcitedefaultmidpunct}
{\mcitedefaultendpunct}{\mcitedefaultseppunct}\relax
\EndOfBibitem
\bibitem[Fest(2013)]{Fest_2013}
Fest,~E. \emph{Stray Light Analysis and Control}; SPIE, 2013\relax
\mciteBstWouldAddEndPuncttrue
\mciteSetBstMidEndSepPunct{\mcitedefaultmidpunct}
{\mcitedefaultendpunct}{\mcitedefaultseppunct}\relax
\EndOfBibitem
\bibitem[Clapham and Hutley(1973)Clapham, and Hutley]{Clapham_1973}
Clapham,~P.; Hutley,~M. Reduction of lens reflexion by the moth eye principle.
  \emph{Nature} \textbf{1973}, \emph{244}, 281--282\relax
\mciteBstWouldAddEndPuncttrue
\mciteSetBstMidEndSepPunct{\mcitedefaultmidpunct}
{\mcitedefaultendpunct}{\mcitedefaultseppunct}\relax
\EndOfBibitem
\bibitem[Deparis \latin{et~al.}(2009)Deparis, Vigneron, Agustsson, and
  Decroupet]{Deparis_2009}
Deparis,~O.; Vigneron,~J.-P.; Agustsson,~O.; Decroupet,~D. Optimization of
  photonics for corrugated thin-films solar cells. \emph{J. Appl. Phys.}
  \textbf{2009}, \emph{106}, 094505\relax
\mciteBstWouldAddEndPuncttrue
\mciteSetBstMidEndSepPunct{\mcitedefaultmidpunct}
{\mcitedefaultendpunct}{\mcitedefaultseppunct}\relax
\EndOfBibitem
\bibitem[Prodan \latin{et~al.}(2003)Prodan, Radloff, Halas, and
  Nordlander]{Prodan_2003}
Prodan,~E.; Radloff,~C.; Halas,~N.; Nordlander,~P. A hybridization model for
  the plasmon response of complex nanostructures. \emph{Science} \textbf{2003},
  \emph{302}, 419--422\relax
\mciteBstWouldAddEndPuncttrue
\mciteSetBstMidEndSepPunct{\mcitedefaultmidpunct}
{\mcitedefaultendpunct}{\mcitedefaultseppunct}\relax
\EndOfBibitem
\bibitem[Christ \latin{et~al.}(2006)Christ, Zentgraf, Tikhodeev, Gippius, Kuhl,
  and Giessen]{Christ_2006}
Christ,~A.; Zentgraf,~T.; Tikhodeev,~S.; Gippius,~N.; Kuhl,~J.; Giessen,~H.
  Controlling the interaction between localized and delocalized surface plasmon
  modes: experiment and numerical calculations. \emph{Phys. Rev. B}
  \textbf{2006}, \emph{74}, 155435\relax
\mciteBstWouldAddEndPuncttrue
\mciteSetBstMidEndSepPunct{\mcitedefaultmidpunct}
{\mcitedefaultendpunct}{\mcitedefaultseppunct}\relax
\EndOfBibitem
\bibitem[Liu \latin{et~al.}(2007)Liu, Guo, Fu, Kaiser, Schweizer, and
  Giessen]{Liu_2007}
Liu,~N.; Guo,~H.; Fu,~L.; Kaiser,~S.; Schweizer,~H.; Giessen,~H. Plasmon
  hybridization in stacked cut-wire metamaterials. \emph{Adv. Mater.}
  \textbf{2007}, \emph{19}, 3628--3632\relax
\mciteBstWouldAddEndPuncttrue
\mciteSetBstMidEndSepPunct{\mcitedefaultmidpunct}
{\mcitedefaultendpunct}{\mcitedefaultseppunct}\relax
\EndOfBibitem
\bibitem[Pu \latin{et~al.}(2012)Pu, Feng, Hu, and Luo]{Pu_2012}
Pu,~M.; Feng,~Q.; Hu,~C.; Luo,~X. Perfect absorption of light by coherently
  induced plasmon hybridization in ultrathin metamaterial film.
  \emph{Plasmonics} \textbf{2012}, \emph{7}, 733--738\relax
\mciteBstWouldAddEndPuncttrue
\mciteSetBstMidEndSepPunct{\mcitedefaultmidpunct}
{\mcitedefaultendpunct}{\mcitedefaultseppunct}\relax
\EndOfBibitem
\bibitem[Lobet \latin{et~al.}(2014)Lobet, Lard, Sarrazin, Deparis, and
  Henrard]{Lobet_2014}
Lobet,~M.; Lard,~M.; Sarrazin,~M.; Deparis,~O.; Henrard,~L. Plasmon
  hybridization in pyramidal metamaterials: a route towards ultra-broadband
  absorption. \emph{Opt. Express} \textbf{2014}, \emph{22}, 12678--12690\relax
\mciteBstWouldAddEndPuncttrue
\mciteSetBstMidEndSepPunct{\mcitedefaultmidpunct}
{\mcitedefaultendpunct}{\mcitedefaultseppunct}\relax
\EndOfBibitem
\bibitem[Lobet and Henrard(2014)Lobet, and Henrard]{Lobet2014bis}
Lobet,~M.; Henrard,~L. Metamaterials for ultra-broadband super absorbers based
  on plasmon hybridization. 8th International Congress on Advanced
  Electromagnetic Materials in Microwaves and Optics. 2014; pp 190--192\relax
\mciteBstWouldAddEndPuncttrue
\mciteSetBstMidEndSepPunct{\mcitedefaultmidpunct}
{\mcitedefaultendpunct}{\mcitedefaultseppunct}\relax
\EndOfBibitem
\bibitem[Beadie \latin{et~al.}(2015)Beadie, Brindza, Flynn, Rosenberg, and
  Shirk]{Beadie_2015}
Beadie,~G.; Brindza,~M.; Flynn,~R.; Rosenberg,~A.; Shirk,~J. Refractive index
  measurements of poly(methyl methacrylate) (PMMA) from 0.4-1.6 $\mu$m.
  \emph{Appl. Optics} \textbf{2015}, \emph{54}, 139\relax
\mciteBstWouldAddEndPuncttrue
\mciteSetBstMidEndSepPunct{\mcitedefaultmidpunct}
{\mcitedefaultendpunct}{\mcitedefaultseppunct}\relax
\EndOfBibitem
\bibitem[Johnson and Christy(1972)Johnson, and Christy]{Johnson_1972}
Johnson,~P.; Christy,~R. Optical constants of noble metals. \emph{Phys. Rev. B}
  \textbf{1972}, \emph{6}, 4370--4379\relax
\mciteBstWouldAddEndPuncttrue
\mciteSetBstMidEndSepPunct{\mcitedefaultmidpunct}
{\mcitedefaultendpunct}{\mcitedefaultseppunct}\relax
\EndOfBibitem
\bibitem[Johnson and Christy(1974)Johnson, and Christy]{Johnson_1974}
Johnson,~P.; Christy,~R. Optical constants of transition metals: Ti, V, Cr, Mn,
  Fe, Co, Ni, and Pd. \emph{Phys. Rev. B} \textbf{1974}, \emph{9},
  5056--5070\relax
\mciteBstWouldAddEndPuncttrue
\mciteSetBstMidEndSepPunct{\mcitedefaultmidpunct}
{\mcitedefaultendpunct}{\mcitedefaultseppunct}\relax
\EndOfBibitem
\bibitem[Rakic(1995)]{Rakic_1995}
Rakic,~A. Algorithm for the determination of intrinsic optical constants of
  metal films: application to aluminum. \emph{Appl. Optics} \textbf{1995},
  \emph{34}, 4755\relax
\mciteBstWouldAddEndPuncttrue
\mciteSetBstMidEndSepPunct{\mcitedefaultmidpunct}
{\mcitedefaultendpunct}{\mcitedefaultseppunct}\relax
\EndOfBibitem
\bibitem[Rakic \latin{et~al.}(1998)Rakic, Djurisic, Elazar, and
  Majewski]{Rakic_1998}
Rakic,~A.; Djurisic,~A.; Elazar,~J.; Majewski,~M. Optical properties of
  metallic films for vertical-cavity optoelectronic devices. \emph{Appl.
  Optics} \textbf{1998}, \emph{37}, 5271--5283\relax
\mciteBstWouldAddEndPuncttrue
\mciteSetBstMidEndSepPunct{\mcitedefaultmidpunct}
{\mcitedefaultendpunct}{\mcitedefaultseppunct}\relax
\EndOfBibitem
\bibitem[Dennis~Jr. and Schnabel(1996)Dennis~Jr., and Schnabel]{Dennis_1996}
Dennis~Jr.,~J.; Schnabel,~R.~B. \emph{Numerical Methods for Unconstrained
  Optimization and Nonlinear Equations}; SIAM: Philadelphia, 1996\relax
\mciteBstWouldAddEndPuncttrue
\mciteSetBstMidEndSepPunct{\mcitedefaultmidpunct}
{\mcitedefaultendpunct}{\mcitedefaultseppunct}\relax
\EndOfBibitem
\bibitem[Goldberg(1989)]{Goldberg_1989}
Goldberg,~D. \emph{Genetic Algorithms in Search, Optimization and Machine
  Learning}; Addison-Wesley: Reading, Mass., 1989\relax
\mciteBstWouldAddEndPuncttrue
\mciteSetBstMidEndSepPunct{\mcitedefaultmidpunct}
{\mcitedefaultendpunct}{\mcitedefaultseppunct}\relax
\EndOfBibitem
\bibitem[Haupt and Werner(2007)Haupt, and Werner]{Haupt_2007}
Haupt,~R.; Werner,~D. \emph{Genetic Algorithms in Electromagnetics}; J. Wiley
  \& Sons: Hoboken, NJ, 2007\relax
\mciteBstWouldAddEndPuncttrue
\mciteSetBstMidEndSepPunct{\mcitedefaultmidpunct}
{\mcitedefaultendpunct}{\mcitedefaultseppunct}\relax
\EndOfBibitem
\bibitem[Eiben and Smith(2007)Eiben, and Smith]{Eiben_2007}
Eiben,~A.; Smith,~J. \emph{Introduction to Evolutionary Computing}, 2nd ed.;
  Springer-Verlag: Berlin, 2007\relax
\mciteBstWouldAddEndPuncttrue
\mciteSetBstMidEndSepPunct{\mcitedefaultmidpunct}
{\mcitedefaultendpunct}{\mcitedefaultseppunct}\relax
\EndOfBibitem
\bibitem[Kennedy and Eberhart(1995)Kennedy, and Eberhart]{Kennedy_1995}
Kennedy,~J.; Eberhart,~R. Particle Swarm Optimization. \emph{Proceedings of
  ICNN'95 - International Conference on Neural Networks} \textbf{1995},
  \emph{4}, 1942--1948\relax
\mciteBstWouldAddEndPuncttrue
\mciteSetBstMidEndSepPunct{\mcitedefaultmidpunct}
{\mcitedefaultendpunct}{\mcitedefaultseppunct}\relax
\EndOfBibitem
\bibitem[Shi and Eberhart(1998)Shi, and Eberhart]{Shi_1998}
Shi,~Y.; Eberhart,~R. A Modified Particle Swarm Optimizer. \emph{Proceedings of
  IEEE International Conference on Evolutionary Computation} \textbf{1998},
  69--73\relax
\mciteBstWouldAddEndPuncttrue
\mciteSetBstMidEndSepPunct{\mcitedefaultmidpunct}
{\mcitedefaultendpunct}{\mcitedefaultseppunct}\relax
\EndOfBibitem
\bibitem[Bonyadi and Michalewicz(2017)Bonyadi, and Michalewicz]{Bonyadi_2017}
Bonyadi,~M.; Michalewicz,~Z. Particle swarm optimization for single objective
  continuous space problems: a review. \emph{Evolutionary Computation}
  \textbf{2017}, \emph{25}, 1--54\relax
\mciteBstWouldAddEndPuncttrue
\mciteSetBstMidEndSepPunct{\mcitedefaultmidpunct}
{\mcitedefaultendpunct}{\mcitedefaultseppunct}\relax
\EndOfBibitem
\bibitem[Dorigo and Stützle(2004)Dorigo, and Stützle]{Dorigo_2004}
Dorigo,~M.; Stützle,~T. \emph{Ant Colony Optimization}; MIT Press: Cambridge,
  MI, 2004\relax
\mciteBstWouldAddEndPuncttrue
\mciteSetBstMidEndSepPunct{\mcitedefaultmidpunct}
{\mcitedefaultendpunct}{\mcitedefaultseppunct}\relax
\EndOfBibitem
\bibitem[Gaier \latin{et~al.}(2020)Gaier, Asteroth, and Mouret]{Gaier_2020}
Gaier,~A.; Asteroth,~A.; Mouret,~J.-B. Discovering Representations for
  Black-box Optimization. \emph{Proceedings of the Genetic and Evolutionary
  Computation Conference (GECCO ’20)} \textbf{2020}, 103--111\relax
\mciteBstWouldAddEndPuncttrue
\mciteSetBstMidEndSepPunct{\mcitedefaultmidpunct}
{\mcitedefaultendpunct}{\mcitedefaultseppunct}\relax
\EndOfBibitem
\bibitem[Moharam and Gaylord(1981)Moharam, and Gaylord]{Moharam_1981}
Moharam,~M.; Gaylord,~T. Rigorous coupled-wave analysis of planar-grating
  diffraction. \emph{J. Opt. Soc. Am. A} \textbf{1981}, \emph{71},
  811--818\relax
\mciteBstWouldAddEndPuncttrue
\mciteSetBstMidEndSepPunct{\mcitedefaultmidpunct}
{\mcitedefaultendpunct}{\mcitedefaultseppunct}\relax
\EndOfBibitem
\bibitem[Khurgin(2017)]{Khurgin_2017}
Khurgin,~J. Replacing noble metals with alternative materials in plasmonics and
  metamaterials: how good an idea? \emph{Phil. Trans. R. Soc. A} \textbf{2017},
  \emph{375}, 20160068\relax
\mciteBstWouldAddEndPuncttrue
\mciteSetBstMidEndSepPunct{\mcitedefaultmidpunct}
{\mcitedefaultendpunct}{\mcitedefaultseppunct}\relax
\EndOfBibitem
\bibitem[Brenner(2010)]{Brenner_2010}
Brenner,~K.-H. Aspects for calculating local absorption with the rigorous
  coupled-wave method. \emph{Opt. Express} \textbf{2010}, \emph{18},
  10369--10376\relax
\mciteBstWouldAddEndPuncttrue
\mciteSetBstMidEndSepPunct{\mcitedefaultmidpunct}
{\mcitedefaultendpunct}{\mcitedefaultseppunct}\relax
\EndOfBibitem
\end{mcitethebibliography}

\end{document}